\documentclass[acmsmall,screen,nonacm]{acmart}

\usepackage{amsmath}
\usepackage{algorithmic}
\usepackage{graphicx}
\usepackage{textcomp}
\usepackage{xcolor}

\usepackage{subcaption}
\usepackage{booktabs}
\usepackage{xspace}
\usepackage{tcolorbox}
\usepackage{multirow}
\usepackage{placeins}

\tcbset{before skip=6pt, after skip=6pt}

\newcommand{\dataset}[1]{\textsc{#1}\xspace}

\newcommand{\RANDOM}{\textsc{Random}\xspace}
\newcommand{\FaRM}{\textsc{FaRM}\xspace}
\newcommand{\Sub}{\textsc{Subsuming}\xspace}

\newcommand{\LinePool}{\textsc{LinePool}\xspace}
\newcommand{\LP}[2]{\textsc{LinePool}(#1, #2)}

\newcommand{\FaRMLine}{\FaRM+\textsc{Line}\xspace}
\newcommand{\SubLine}{\Sub+\textsc{Line}\xspace}

\newcommand{\RandSig}{\textsc{Rand}\xspace}

\begin{document}

\title{Where Does the Budget Go? Structural Concentration in Learning-Based Mutant Selection}

\author{Taha Rostami}
\affiliation{%
  \institution{University of Luxembourg}
  \city{Luxembourg}
  \country{Luxembourg}}
\email{taha.rostami@uni.lu}

\author{Jeongju Sohn}
\affiliation{%
  \institution{Kyungpook National University}
  \city{Daegu}
  \country{Republic of Korea}}
\email{jeongju.sohn@knu.ac.kr}

\author{Mike Papadakis}
\affiliation{%
  \institution{University of Luxembourg}
  \city{Luxembourg}
  \country{Luxembourg}}
\email{michail.papadakis@uni.lu}

\begin{abstract}
Learning-based mutant selection is used to reduce the cost of mutation testing by ranking and selecting the mutants that seem to be the most promising. Prior work has shown great potential for this approach in fault revelation and subsuming-mutant selection. However, the risks and benefits of these approaches---especially the distribution of selected mutants across code locations and how that distribution affects behavioral diversity and fault revelation---remain unclear.
We investigate this distribution and its effects, and introduce \LinePool, a simple, model-agnostic stratification step that distributes selections across source lines while retaining the underlying ranking signal. 
Across two datasets and budgets (2\% and 5\% of the mutants), we find that learning-based selection concentrates the budget on a few code locations, and may thereby systematically introduce `blind spots', untested code areas, allowing faults to escape detection. On larger programs, the selected mutants also exhibit substantial overlap in kill behavior. 
\LinePool substantially increases spatial coverage and reduces kill-set overlap. It also improves fault revelation at tight budgets and makes selection more robust to imperfect or shifted ranking signals.
Comparisons with clustering and established diversification methods show that none consistently outperforms \LinePool across datasets, supporting its use as a simple diversification step. 
In general, our results suggest that behavioral diversity should be considered alongside effectiveness when designing and evaluating mutant-selection methods, and that exploiting structural elements of the program is one simple way to obtain such diversity while maintaining effectiveness.
\end{abstract}
\maketitle

\section{Introduction}

Mutation testing assesses and improves test suites by introducing simple syntactic changes (mutants) and checking whether tests detect the resulting behavioral differences. Although highly effective, its practical use is limited by cost: mutation tools generate many mutants, and executing them is computationally expensive \cite{PapadakisEtAl2019AdvancesInComputersSurvey}.

A common response is to \emph{select} a budgeted subset of mutants. By reducing the number of the considered mutants one can reduce the overall cost associated with mutation testing since fewer mutants need to be generated, executed and analyzed. Recent work increasingly relies on learning-based ranking, which predicts a utility score for each mutant and selects the top-$k$ mutants under a fixed budget ~\cite{TitcheuChekamEtAl2020SelectingFaultRevealing, GargEtAl2023Cerebro}. Such approaches provide favorable cost–effectiveness tradeoffs.

However, these approaches are evaluated almost exclusively through their effectiveness, while the diversity of the selected mutant set is rarely examined. Under a tight budget, this matters: many mutants share code context and operator patterns, which may cause prediction models to assign high scores to mutants from the same source region, concentrating the selection budget on only a few lines.

Importantly, structural concentration is not inherently undesirable---a ``hot'' region may legitimately deserve additional attention. The concern is what a concentrated budget leaves out and what it spends twice. First, concentration may introduce \emph{blind spots}: code regions, possibly containing the fault, for which no mutant is selected. Second, it may lead to \emph{behavioral redundancy}: when many selected mutants exhibit overlapping kill behavior, a large fraction of the budget is spent on redundant testing signals. Distribution shift can further amplify both effects.

We study two representative ranking signals: FaRM, a fault-revealing mutant predictor \cite{TitcheuChekamEtAl2020SelectingFaultRevealing}, and a subsuming-mutant predictor (\Sub) inspired by prior work \cite{GargEtAl2023Cerebro}. We evaluate them on mutation datasets derived from the \dataset{Codeflaws}~\cite{CodeFlaws2017icse-c} and \dataset{CoREBench}~\cite{CoreBench2014issta} benchmarks.

We show that score-based top-$k$ selection often produces structural concentration with both consequences. On \dataset{Codeflaws}, top-$k$ selection omits the faulty line in about half of the subjects, and the fault-revelation gains from diversifying selection are concentrated in these cases. On the larger \dataset{CoREBench} programs, concentration is additionally coupled with behavioral redundancy, and unpooled selection can reveal fewer faults than random selection.

To address this problem, we introduce \LinePool, a simple model-agnostic post-processing strategy that groups mutants by source line and uses the original ranking signal to prioritize both lines and mutants within each line. It requires no additional mutant features or parameter tuning, making it straightforward to apply to an existing ranking method. By spreading the selection across source lines, \LinePool limits spatial blind spots and behavioral overlap while retaining the original ranking signal. To reflect the growing interest in Large Language Model (LLM)-based approaches, we further examine LLM-based ranking and find evidence of the same failure mode, with \LinePool mitigating it in the same way.

The contributions of this paper are:
\begin{itemize}
    \item \textbf{Diagnosis of structural concentration and its consequences.} We empirically show that score-based top-$k$ selection can collapse structurally, leaving blind spots that frequently omit the faulty line, and that, particularly on larger programs, this collapse is coupled with behavioral redundancy. 
    
\item \textbf{A lightweight, model-agnostic selection operator.} We propose \LinePool, a rank-preserving diversification operator that can be applied on top of any mutant-ranking signal. It reallocates a fixed mutation budget across source locations while preserving within-location ranking, thereby limiting blind spots and reducing behavioral overlap.
    
    \item \textbf{Large-scale evaluation.} 
    On mutation datasets derived from \dataset{Codeflaws} (1{,}692 programs; 1.76M mutants) and \dataset{CoREBench} (45 regression faults)~\cite{TitcheuChekamEtAl2020SelectingFaultRevealing}, \LinePool consistently increases spatial coverage, reduces overlap, and preserves ranking effectiveness---improving fault revelation at tight budgets. Comparisons with clustering and established diversification methods show that none consistently outperforms \LinePool, while \LinePool remains simpler to integrate in practice.
\end{itemize}

\section{Background and Related Work}
\label{sec:background}
\subsection{Mutation Testing}

Mutation testing is commonly used for assessing the effectiveness of test suites \cite{PapadakisEtAl2019AdvancesInComputersSurvey,PizzoletoEtAl2019SLRMutationCost}. It introduces small syntactic changes, called \emph{mutations}, into the original program to produce modified programs known as \emph{mutants}. Each mutant represents a plausible fault, such as replacing an operator, modifying a constant, or altering a conditional expression. The test suite is executed against each mutant. If at least one test distinguishes its behavior from that of the original program, the mutant is said to be \emph{killed}; otherwise, it remains \emph{live}.

The mutation score---the proportion of killed mutants---is a strong indicator of test-suite quality and has been shown to correlate well with real fault detection \cite{GaySalahirad2023CoupledOperators}. 
Beyond evaluation, mutation testing has been used to guide automated test generation \cite{FosterEtAl2025MetaMutationGuidedLLM,ChekamEtAl2021StubbornMutants}, where surviving mutants highlight inadequately exercised program behaviors and guide the generation of tests to kill them.

\subsection{Mutant Selection}

Given the high computational cost of mutation testing, a substantial body of research has focused on reducing the number of mutants while preserving effectiveness.

One early direction is \emph{selective mutation}, which restricts the applied mutation operators. Mathur \cite{mathur1991performance} introduced constrained mutation using a small subset of operators, and subsequent studies by Wong and Mathur \cite{WongMathur1995CostReduction} and Offutt et al.~\cite{OffuttRothermelZapf1993SelectiveMutation} demonstrated that carefully chosen operator subsets can significantly reduce mutant counts with minimal loss in effectiveness. These reduced operator sets have since been widely adopted by mutation testing tools \cite{PapadakisEtAl2019AdvancesInComputersSurvey}.

Other approaches exploit program structure or mutant similarity. Clustering-based techniques, for example, group mutants according to similarity and select representative mutants from the resulting clusters \cite{Hussain2008MutationClustering}. Later work explored structural properties of the program, including mutation location and control-flow relationships. Dominator-based techniques prioritize mutants whose execution or coverage can imply that of others \cite{SunEtAl2017PathAwareReduction,GongEtAl2017DominanceWeakMutation}.

Despite these efforts, extensive empirical evidence suggests that many reduction techniques offer limited improvement over random selection \cite{GopinathEtAl2016LimitsReduction}.
Prior studies have shown that random sampling often achieves effectiveness comparable to more complex heuristics, and that many mutants are weakly related to real faults \cite{LaurentEtAl2022Defects4J2Coupling, PapadakisEtAl2018ICSECorrelation, GargEtAl2024Vul4JLLMMutants}.

\subsection{Learning-Based Mutant Selection}\label{sub:learning_based_mutant_selection}

Recent work has shifted toward targeting \emph{useful} mutants using supervised learning to predict those more likely to reveal real faults or subsume others, rather than relying solely on handcrafted heuristics \cite{TitcheuChekamEtAl2020SelectingFaultRevealing, GargEtAl2023Cerebro, RostamiJalili2023FrMi}.

Titcheu Chekam et al.~\cite{TitcheuChekamEtAl2020SelectingFaultRevealing} introduced \FaRM, a learning-based approach that extracts static features from mutants and trains classifiers to predict fault-revealing mutants. A mutant is \emph{fault-revealing} if it is killable and the tests that kill it also reveal real faults in the program.
Mutants are then prioritized according to predicted usefulness, and the top-ranked subset is selected for further analysis or test generation.
Experimental results demonstrated improvements in fault revelation compared to traditional selection methods, including random sampling and operator-based selection.

More recently, Garg et al.~\cite{GargEtAl2023Cerebro} proposed Cerebro, a neural encoder--decoder model designed to identify subsuming mutants based on mutation operators and code context. Subsuming mutants are mutants whose kill sets subsume those of others, such that killing a subsuming mutant implies killing one or more dominated mutants.

While prior work focuses primarily on improving fault revelation effectiveness,
the properties of the resulting mutant sets, such as spatial
concentration and behavioral redundancy, remain largely unexplored.
Our work complements existing learning-based selection approaches by
examining these effects and showing how they influence the
diversity and robustness of the selected mutants.

\section{Mutant Selection Strategies}
\label{sec:mutant_sel_strategies}
Given a program $P$ and a mutant set $M$, a mutant selection approach chooses a subset $S \subseteq M$ of size $k$ to optimize a target objective, such as fault revelation.

The simplest strategy is \RANDOM, which selects mutants uniformly at random.
Despite its simplicity, random selection has shown to be a strong baseline~\cite{GopinathEtAl2016LimitsReduction}.

More recently, learning-based approaches have been proposed to predict mutant utility.
Given a predicted usefulness score $s(m)$ for each mutant $m \in M$, these approaches select the top-$k$ ranked mutants according to $s(m)$. We study two learning-based strategies:

\begin{itemize}
    \item \textbf{Fault-revealing ranking} (\FaRM), which prioritizes mutants predicted to be fault-revealing.
    \item \textbf{Subsuming ranking} (\Sub), which prioritizes mutants predicted to be subsuming.
\end{itemize}

While top-$k$ selection maximizes predicted usefulness, it may over-concentrate selections if many high-scoring mutants originate from the same source locations, potentially leaving other locations without any selected mutant (blind spots) and inducing behavioral redundancy.

\subsection{Line-Level Pooling}
\label{sec:linepool}

To mitigate the structural concentration induced by top-$k$ selection, we present \LinePool, a model-agnostic post-processing procedure that changes how a ranking is consumed rather than how the ranking is produced. Given any mutant-level ranking, \LinePool first groups mutants by a structural unit, such as source line, and then performs round-based selection across groups. This makes \LinePool a rank-preserving diversification operator: it prevents the global top-$k$ from over-allocating the budget to a few dense locations, while still using the original ranking to prioritize both groups and mutants within each group, thereby promoting \emph{structural diversity} and limiting potential blind spots.

We define the grouping and selection procedure as follows.
Let $l : M \rightarrow \mathcal{L}$ be a grouping function that assigns each mutant to a partition identifier.
For line-level pooling, $l(m)$ returns the source code line number of mutant $m$. We partition $M$ into groups $G = \{G_1, \dots, G_p\}$, where
$G_i = \{\, m \in M \mid l(m) = \ell_i \,\}$ for each distinct identifier $\ell_i \in \mathcal{L}$.

Line-level pooling uses two score functions.
$s_L(m)$ is a mutant-level score used to determine the priority of the \emph{group} containing $m$, and
$s_W(m)$ determines the priority of $m$ within its group.
Group priorities are computed by aggregating $s_L(m)$ over mutants in each group; in our setting, we use the maximum score, i.e., $\max_{m \in G_i} s_L(m)$, and examine alternative aggregation rules in Section~\ref{sec:rq4-alternatives}.

Groups are sorted in descending order of their computed priority, and mutants within each group are sorted in descending order of $s_W(m)$. The final subset $S$ of size $k$ is constructed by iterating over these groups in this order. We first select the highest-ranked mutant (according to $s_W$) from each group. If $|S| < k$, we then select the second-highest mutant from each group in the same way, and continue until $|S| = k$ or all groups are exhausted. This prioritizes selecting from different groups before allocating multiple mutants to the same group, while preserving the within-group ranking induced by $s_W$.

\LinePool decouples structural budget allocation from mutant ranking. The group-level signal 
$s_L$ determines where the mutation-testing budget is spent, i.e., which source locations are prioritized, while the within-group signal $s_W$ determines which mutants are selected within each location. Consequently, \LinePool is applicable on top of existing learning-based or LLM-based ranking approaches without retraining or modifying the underlying model.

In the main evaluation, we set $s_L = s_W$, using either \FaRM or \Sub as the single scoring signal for both group prioritization and within-group ranking, resulting in two configurations: \FaRMLine and \SubLine.
Under this setting, group priorities are determined by the highest-ranked mutant within each group.

\section{Experimental Setup}
\label{sec:setup}
\subsection{Datasets}

We use publicly available mutation datasets constructed by prior work~\cite{TitcheuChekamEtAl2020SelectingFaultRevealing}, from two C benchmarks: \dataset{Codeflaws}~\cite{CodeFlaws2017icse-c} and \dataset{CoREBench}~\cite{CoreBench2014issta}; these datasets were obtained by generating mutants and filtering them to manage mutation testing cost. 
For simplicity, we refer to these derived mutation datasets as \dataset{Codeflaws} and \dataset{CoREBench}, respectively.   

\paragraph{\dataset{Codeflaws}.}
\dataset{Codeflaws} contains 1,692 C programs collected from online programming competitions (1 to 322 LOC; average 36 LOC), with 1,756,031 generated mutants (over 1,000 mutants per program on average).
Developer tests are augmented with automatically generated tests (KLEE \cite{cadar2008klee}) using extended time limits.

\paragraph{\dataset{CoREBench}.}
\dataset{CoREBench} contains 45 real regression faults from four large C projects, enabling evaluation on realistic regression scenarios beyond small benchmark programs. 
Developer tests are augmented with automatically generated tests (KLEE \cite{cadar2008klee}) using extended time limits.

\paragraph{Dataset selection and scope.}
Both datasets provide complete mutant-test kill information alongside real faults. Such information is required to evaluate mutant selection for fault revelation. Few publicly available datasets provide it, and generating complete mutant-test kill information is costly: constructing the \dataset{Codeflaws} mutation dataset alone required 8,009 CPU days of mutation analysis~\cite{TitcheuChekamEtAl2020SelectingFaultRevealing}. 
\dataset{Codeflaws} offers many subjects, but its programs are small and do not reproduce the context of large production systems.  
In practice, small mutation scopes nevertheless occur: mutation analysis often focuses on code changed during review~\cite{Petrovic2018Google} or on mutants relevant to an individual commit~\cite{Ma2021MuDelta}, leaving a limited mutation scope even within a large system. 
\dataset{Codeflaws} 
allows us study selection over a limited amount of code, but not the interactions between the changed code and the surrounding system. 
\dataset{CoREBench} provides a complementary setting with substantially larger programs and regression faults. 
We discuss the implications of both datasets for generalizability in Section~\ref{sec:threats}.  

\subsection{Learning-Based Models}
\label{sec:models}

Section~\ref{sec:mutant_sel_strategies} introduces two learning-based mutant selection strategies: 1) \emph{fault-revealing mutants (\FaRM)} and 2) \emph{subsuming mutants (\Sub)}. For each strategy, we use existing models from prior work to obtain ranking scores of mutants. 

\textbf{For the \FaRM strategy}, we reuse the fault-revealing mutant prediction model proposed by Titcheu Chekam et al.~\cite{TitcheuChekamEtAl2020SelectingFaultRevealing} and adopted in the Mart mutation testing framework \cite{ChekamEtAl2019MART}. The model assigns a real-valued usefulness score to each mutant using static features extracted from the mutant and its surrounding code context. In our study, we treat these scores purely as \emph{a ranking signal}: we do \emph{not} retrain or recalibrate the model, and we directly apply score-based top-$k$ selection (and pooled variants) using the scores output by the released implementation.

\textbf{Regarding \Sub strategy}, Garg et al.~\cite{GargEtAl2023Cerebro} proposed Cerebro as well as a feature-based baseline trained on static mutant features. We adopt this \emph{feature-based baseline} and train a gradient-boosted decision tree model~\cite{Friedman2002StochasticGB} (1{,}000 trees, maximum depth 5; settings as in~\cite{GargEtAl2023Cerebro}). We use 10-fold cross-validation and aggregate out-of-fold predictions to obtain an unbiased subsuming score for every mutant. We denote this ranking signal as \Sub and use it purely as a ranking signal for selection under fixed budgets, including the line-level pooling variants.

We do not evaluate Cerebro's neural model, as it relies on mutation-aware code representations (e.g., exact location annotations) that are not directly compatible with the mutant format provided by \dataset{Codeflaws}/\dataset{CoREBench}. Adapting this pipeline would require reconstructing such representations and introducing additional assumptions about mutation-to-source alignment, which may affect comparability. 

\subsection{Evaluation Metrics}

We assess structural concentration, blind spots, behavioral redundancy, effectiveness, cost, and robustness under fixed selection budgets of 2\% and 5\%, consistent with prior work \cite{TitcheuChekamEtAl2020SelectingFaultRevealing}. We focus on these relatively small budgets because they are the most informative for distinguishing selection strategies: as the budget increases, outcomes naturally tend to converge, whereas substantially smaller budgets can become particularly restrictive for \dataset{Codeflaws}, where the absolute number of selected mutants may be very small. We additionally evaluated budgets of 0.5\%, 1\%, 10\%, and 20\%: the advantage of \LinePool over unpooled top-$k$ selection is largest at the tightest budgets and diminishes as the budget grows, but the conclusions of the paper remain unchanged, so we omit these results for brevity.

\subsubsection{Structural concentration}
We report:
(i) \textbf{LineCov}: the proportion of distinct source lines covered by the selected mutants relative to all mutant-bearing lines (higher is better),
(ii) \textbf{MaxShare}: the maximum fraction of selected mutants from a single line (lower is better),
and (iii) \textbf{Top5Share}: the fraction of selected mutants originating from the five most frequent lines (lower is better). To assess whether concentration leaves consequential regions untested (blind spots), we also report the \textbf{faulty-line omission rate}: the proportion of subjects for which no selected mutant lies on the faulty line (lower is better).

\subsubsection{Behavioral redundancy.}
We quantify behavioral overlap using average pairwise Jaccard similarity \cite{Jaccard1901} between kill sets as suggested by previous studies \cite{OjdanicGKDPT23}.
For mutants $m_i,m_j$ with kill sets $K_i,K_j$:$J(m_i,m_j) = \frac{|K_i \cap K_j|}{|K_i \cup K_j|}$. We report the average pairwise similarity across all selected mutant
pairs for each strategy. Lower values reflect less behavioral overlap
and are preferable.

Redundancy among mutants has also been studied through minimal mutant set analyses~\cite{Ammann2014MinimalMutants}, which aim to identify a reduced set that preserves mutation score. In contrast, Jaccard similarity directly measures behavioral overlap among selected mutants and can therefore distinguish mutant sets with identical mutation adequacy but different levels of behavioral diversity. As complementary sanity checks closely related to minimal mutant sets, we also report the duplication rate, i.e., the proportion of mutants with identical killing behavior, and the subsuming mutation score.

\subsubsection{Fault detection effectiveness.}\label{subsub:fault_detection}
We measure effectiveness using the \emph{Fault Revelation Rate (FRR)}, following established procedures from prior work. In each trial, a mutant is randomly selected, and a test case that kills this mutant is randomly chosen. If the test case reveals a fault, the trial ends. Otherwise, all mutants killed by this test are then removed, and the process repeats until no additional mutants can be killed. Mutants that are not killed by any test are treated as equivalent. We repeat this process 100 times and compute the average probability of revealing real faults. Also, consistent with prior work \cite{TitcheuChekamEtAl2020SelectingFaultRevealing, GargEtAl2023Cerebro}, we compare fault detection effectiveness between approaches using the Wilcoxon signed-rank test \cite{Wilcoxon1945} with Holm correction for multiple comparisons~\cite{Holm1979}.

\subsubsection{Robustness.}\label{subsub:robustness}
In practice, predicted usefulness scores may be imperfect due to model uncertainty, feature approximation, or distribution shift. 
We therefore evaluate robustness under two complementary settings. Unless stated otherwise, these experiments use the 2\% budget.

\textbf{(i) Robustness to ranking noise.}
To assess sensitivity to imperfect ranking scores, for each mutant $m$ with original score $s(m)$ (from \FaRM or \Sub), we perturb the score by adding zero-mean Gaussian noise \cite{hendrycks2019robustness}:
$s_{\lambda}(m) = s(m) + \lambda \cdot \epsilon, \quad \epsilon \sim \mathcal{N}(0,1)$,
where $\lambda$ controls the noise magnitude. We vary $\lambda$ from 0 (no noise) to 1.0 (strong corruption of the ranking signal). For each $\lambda$, we generate multiple independent noisy score realizations, re-run selection using $s_{\lambda}(\cdot)$, and compute FRR. We report the mean FRR across realizations.

\textbf{(ii) Out-of-distribution evaluation.}
To assess out-of-distribution robustness, we depart from the within-benchmark model setup described in Section 4.2: we train both ranking models (\FaRM and \Sub) on one benchmark and evaluate them on the other without using data from the target benchmark for training. This setup mimics deployment scenarios where training and target distributions differ. We evaluate both directions: training on \dataset{Codeflaws} and testing on \dataset{CoREBench}, and vice versa.

\subsubsection{Cost.}\label{subsub:cost}
We treat computational cost as a secondary outcome due to the difficulty of realistic estimation~\cite{TitcheuChekamEtAl2020SelectingFaultRevealing}. We report a simple cost metric primarily as a sanity check, since disproportionate increases may indicate methodological issues. 
We measure cost as the total number of additional program executions during fault revelation: in each round, we execute the selected test case against all $n$ remaining mutants until a real fault is revealed. Since this metric may not fully reflect practical testing effort in isolation, we interpret it together with the number of tests selected until fault revelation. When two approaches require a comparable number of tests, we prefer the one achieving higher FRR, even if its execution-based cost is slightly higher.

\subsubsection{Aggregation.}\label{subsub:aggregation}
Unless stated otherwise, we compute each metric per subject (program for \dataset{Codeflaws}, fault for \dataset{CoREBench}) and then aggregate across subjects.
For \emph{effectiveness} (FRR), each trial is a Bernoulli outcome (fault revealed or not), so a subject's FRR is the empirical success probability over 100 trials; we report the \emph{mean} FRR across subjects, which estimates the expected fault-revelation probability under a uniformly random subject.
For \emph{structural} metrics (LineCov, MaxShare, Top5Share), distributions are often skewed with occasional extreme values, so we report the \emph{median} across subjects to reduce sensitivity to outliers. For \emph{behavioral} overlap (Jaccard), tables report the \emph{mean} across subjects, and figures show the full distributions. For the two sanity-check cost proxies (additional mutant executions and number of selected tests until fault revelation), we likewise report the \emph{median} over all subject--trial runs.

\subsection{Research Questions}
We address the following research questions:
\begin{enumerate}
    \item \textbf{RQ1:} To what extent do mutant selection strategies exhibit structural concentration (e.g., selecting many mutants from few lines), and does this concentration leave blind spots?
    \item \textbf{RQ2:} To what extent do the selected mutants exhibit behavioral redundancy, and how does it relate to structural concentration?
    \item \textbf{RQ3:} How does diversification affect fault revelation effectiveness under fixed selection budgets?
    \item \textbf{RQ4:} What drives the gains of \LinePool: enforcing coverage, preserving within-line ranking, or both?
    \item \textbf{RQ5:} Does line-level pooling improve robustness under distribution shift and noisy or imperfect ranking signals?
\end{enumerate}

\section{Results}
\label{sec:results}

\subsection{RQ1: Structural Concentration and Blind Spots}

\begin{figure*}[t]
\centering
\includegraphics[
    width=0.85\textwidth,
    trim=0cm 0cm 0cm 0cm,
    clip
]{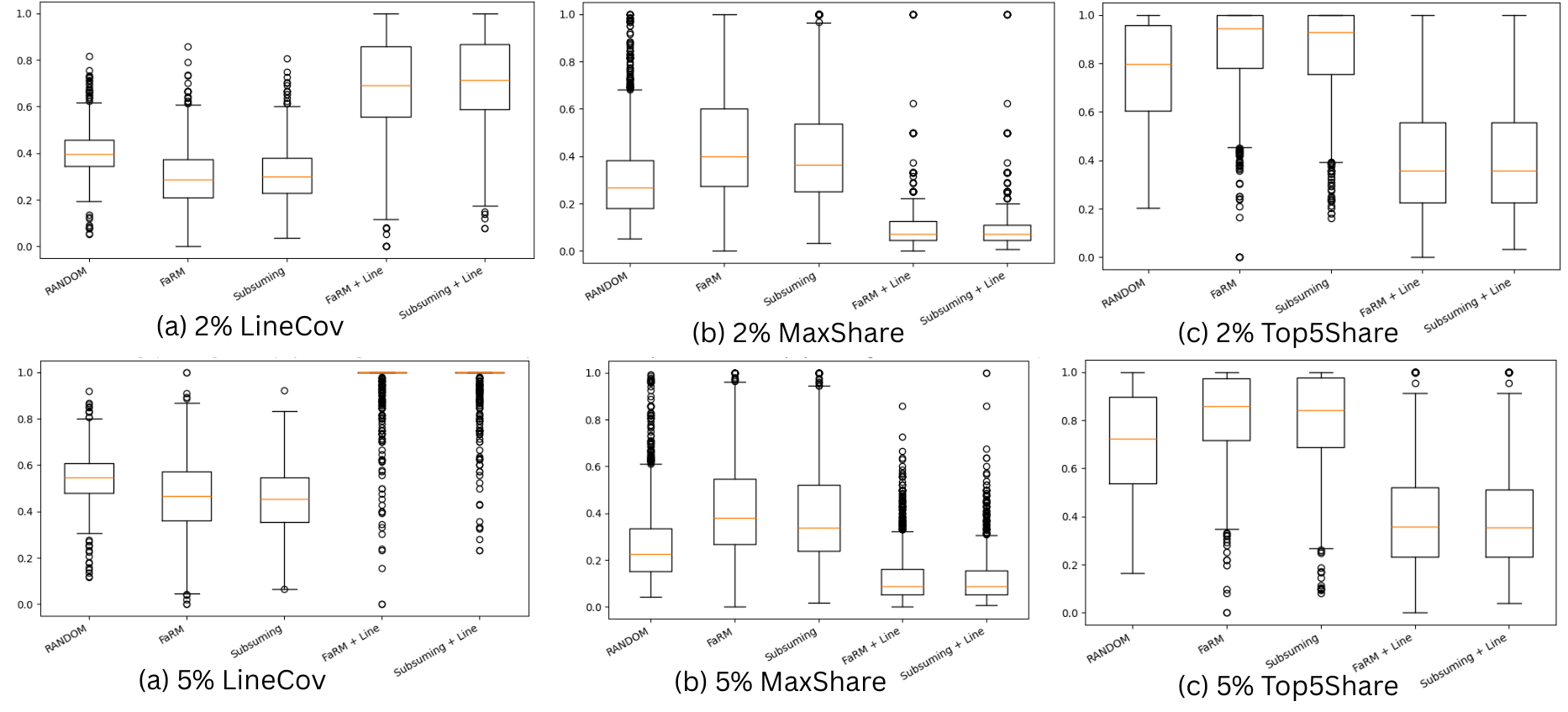}

\caption{Structural concentration (\dataset{Codeflaws}).
LineCov higher-is-better; others lower-is-better. Orange lines are medians.
}
\label{fig:rq1_codeflaws}
\end{figure*}

\begin{figure*}[t]
\centering
\includegraphics[
    width=0.85\textwidth,
    trim=0cm 0cm 0cm 0cm,
    clip
]{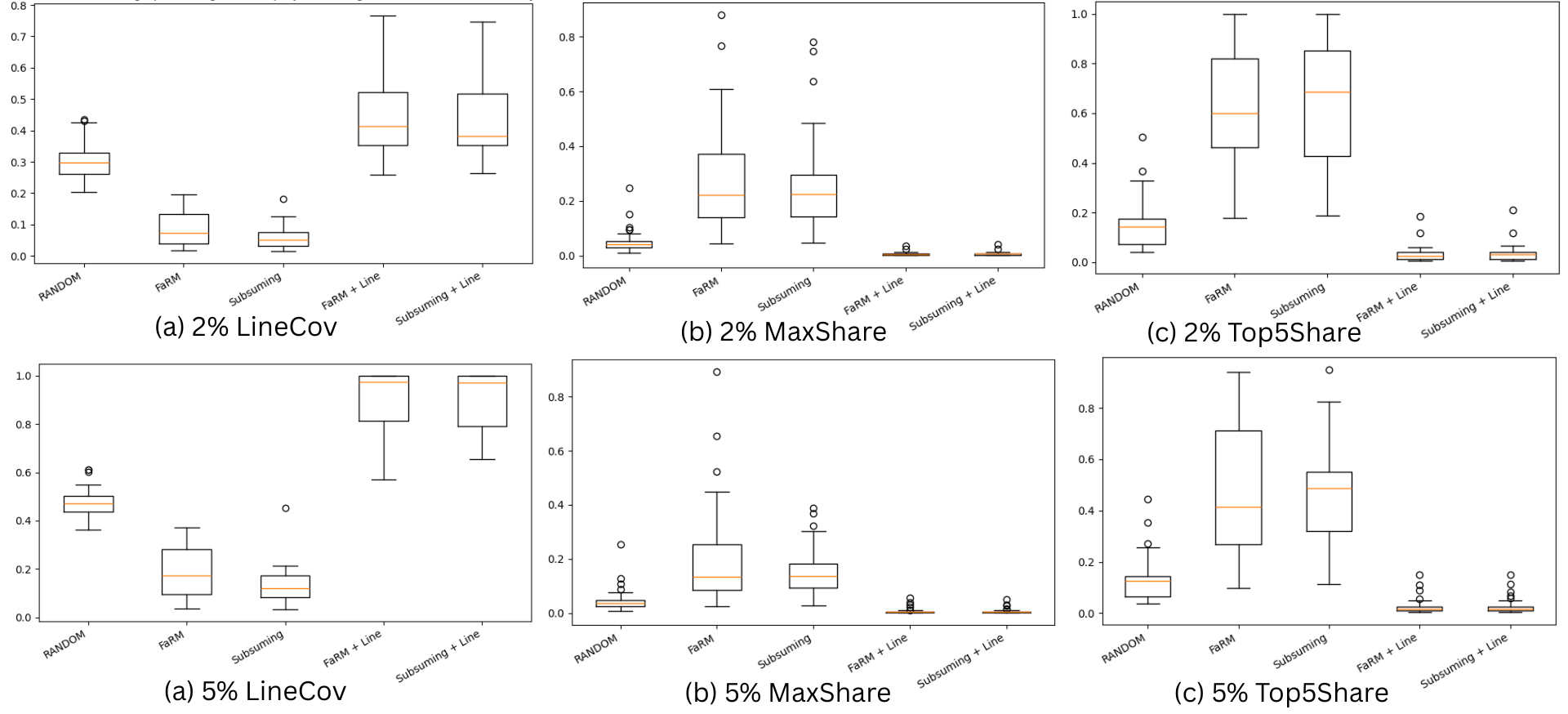}

\caption{Structural concentration on \dataset{CoREBench}. LineCov higher-is-better; others lower-is-better. Orange lines are medians.
}
\label{fig:rq1_corebench}
\end{figure*}

We first examine whether score-based top-$k$ mutant selection concentrates the selection budget on a small number of source lines, i.e., exhibits \emph{structural collapse}. Figures~\ref{fig:rq1_codeflaws} and~\ref{fig:rq1_corebench} show that both learning-based rankers (\FaRM and \Sub) introduce strong \emph{structural concentration}. Compared to \RANDOM, top-$k$ learning-based selection covers substantially fewer mutant-bearing lines (lower LineCov) and allocates a large share of the budget to a handful of locations (higher MaxShare and Top5Share). This happens if a model assigns many relatively high scores to mutants that are syntactically similar or occur near the same code fragment: top-$k$ then repeatedly ``spends'' the budget on the same region.

Line pooling (\FaRMLine, \SubLine) reverses this effect by construction: it forces the selection to represent distinct lines early, lowering MaxShare/Top5Share and increasing LineCov. Notably, pooling does not merely match \RANDOM---it often exceeds it on dispersion, because \RANDOM sample more cases from dense mutant lines, whereas pooling actively counteracts density.

\paragraph{Blind spots: faulty-line coverage.}
To check whether concentration leaves the lines that matter untested, we examine whether the selected mutants include the faulty line on \dataset{Codeflaws}, using the 1{,}464 subjects with a mapped faulty line (see Section~\ref{sec:construct} for \dataset{CoREBench}). At the 2\% budget, \FaRM and \Sub top-$k$ selection omit the faulty line in 53.6\% and 50.8\% of subjects, respectively, whereas \LinePool reduces this to 19.2\% and 15.4\%. In about half of the subjects, top-$k$ selection thus spends the entire budget away from the fault. At 5\%, top-$k$ selection still omits the faulty line in 36.5\% (\FaRM) and 35.2\% (\Sub) of subjects, more often than \RANDOM (28.9\%), whereas \LinePool almost never does (about 1\%).

\begin{tcolorbox}[boxsep=2pt,left=2pt,right=2pt,top=1pt,bottom=1pt]
\noindent\textbf{Answer to RQ1.} 
Learning-based top-$k$ selection strongly concentrates mutants on a small number of lines, leaving blind spots that omit the faulty line in about half of the \dataset{Codeflaws} subjects at 2\% and about a third at 5\%. Line-level pooling restores broad spatial coverage and substantially reduces such omissions.
\end{tcolorbox}

\subsection{RQ2: Behavioral Redundancy}

\begin{figure*}[t]
\centering
\vspace{-0.5em}
\begin{minipage}{0.24\textwidth}
\centering
\includegraphics[width=\linewidth,trim=0.9cm 0cm 0cm 0.85cm,clip]{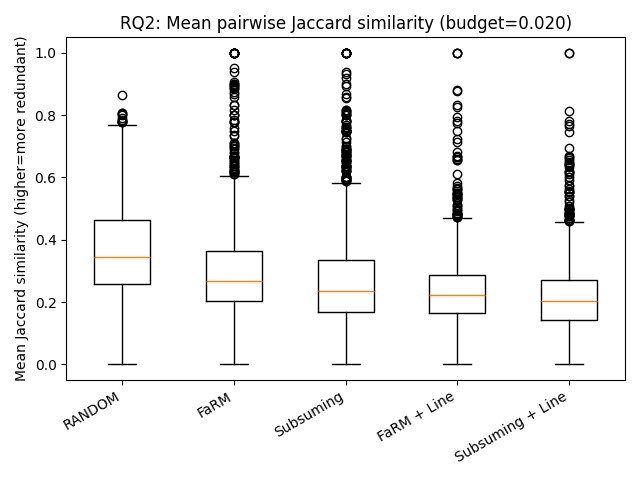}\\
(a) \dataset{Codeflaws} — 2\%
\end{minipage}\hfill
\begin{minipage}{0.24\textwidth}
\centering
\includegraphics[width=\linewidth,trim=0.9cm 0cm 0cm 0.85cm,clip]{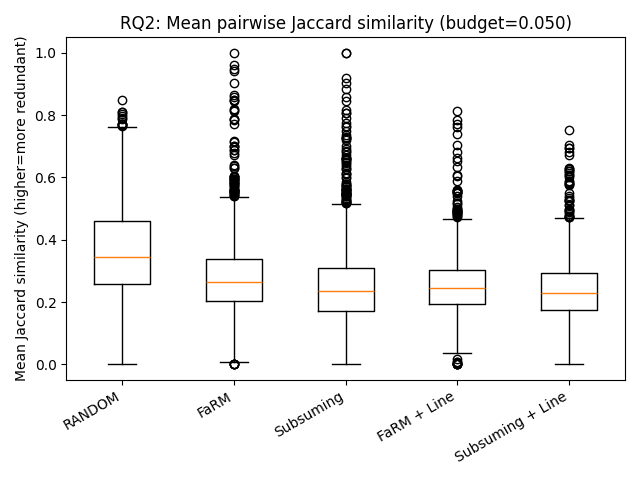}\\
(b) \dataset{Codeflaws} — 5\%
\end{minipage}\hfill
\begin{minipage}{0.24\textwidth}
\centering
\includegraphics[width=\linewidth,trim=0.9cm 0cm 0cm 0.85cm,clip]{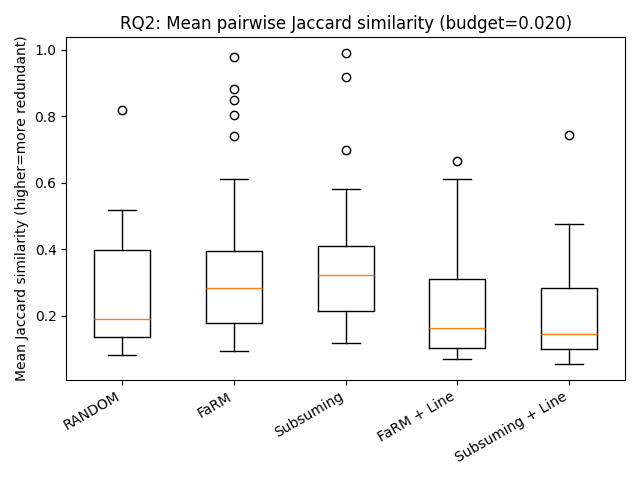}\\
(c) \dataset{CoREBench} — 2\%
\end{minipage}\hfill
\begin{minipage}{0.24\textwidth}
\centering
\includegraphics[width=\linewidth,trim=0.9cm 0cm 0cm 0.85cm,clip]{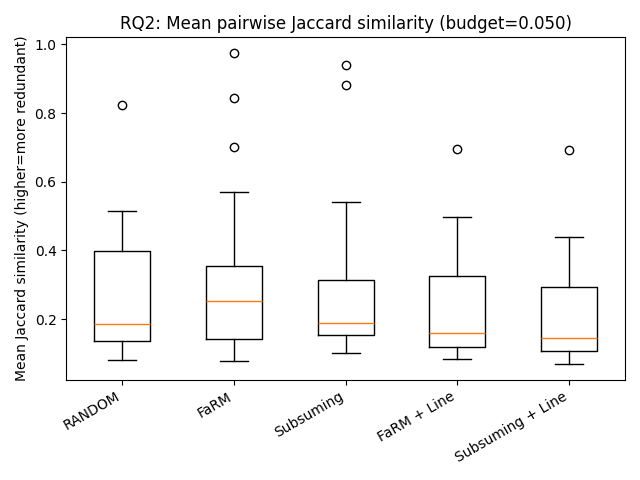}\\
(d) \dataset{CoREBench} — 5\%
\end{minipage}
\vspace{-0.5em}
\caption{Behavioral redundancy measured via average pairwise Jaccard similarity (lower is better). Orange lines are medians.
}
\vspace{-0.5em}
\label{fig:rq2_jaccard}
\end{figure*}

We analyze whether selected mutants exhibit overlapping execution behavior. Figure~\ref{fig:rq2_jaccard} reports pairwise Jaccard similarity between kill sets. The key takeaway is that ranking \emph{does not reliably imply behavioral diversity, though the effect varies with project domain and scale}. On \dataset{Codeflaws}, learning-based ranking tends to reduce overlap relative to \RANDOM, but pooling reduces overlap even further and is consistently the most behaviorally diverse variant. There, top-$k$ selection is more concentrated than \RANDOM yet less redundant, so concentration does not necessarily imply redundancy; its main consequence is the blind spots identified in RQ1. On \dataset{CoREBench}, the effect is more pronounced: at the 2\% budget, unpooled learning-based selection, especially \Sub, yields \emph{higher} redundancy than \RANDOM, in line with RQ1: when selection repeatedly draws from a few lines, many mutants are killed by largely the same tests. Pooling reduces redundancy across both datasets and budgets.

To complement the Jaccard analysis, we performed two additional redundancy analyses at the 2\% budget. First, we measured the reduction in duplicate behavioral profiles among the selected mutants. Across both datasets, \LinePool consistently lowered duplication rates. On \dataset{Codeflaws}, duplication decreased by 11.5 and 13.1 percentage points for \FaRM and \Sub, respectively. The reductions were even larger on \dataset{CoREBench}, reaching 22.9 percentage points for \FaRM and 26.5 percentage points for \Sub. The diversity gains of \LinePool thus translate into substantially fewer redundant mutants.

Second, we evaluated the selected mutant sets against subsuming mutation score. On \dataset{Codeflaws}, \LinePool increased the subsuming mutation score from 86.9\% to 95.9\% for \FaRM and from 85.9\% to 96.2\% for \Sub. On \dataset{CoREBench}, the improvements were even larger, increasing the score from 83.3\% to 100\% for \FaRM and from 62.6\% to 100\% for \Sub. These gains thus also improve the preservation of mutation adequacy under subsumption.

\begin{tcolorbox}[boxsep=2pt,left=2pt,right=2pt,top=1pt,bottom=1pt]
\noindent\textbf{Answer to RQ2.} 
Learning-based top-$k$ selection does not consistently avoid behavioral redundancy: on the larger \dataset{CoREBench} programs, it is more redundant than \RANDOM at the 2\% budget, whereas on \dataset{Codeflaws} its concentration manifests mainly as blind spots. Line pooling reliably reduces redundancy across datasets and budgets.
\end{tcolorbox}

\vspace{-0.8em}

\subsection{RQ3: Fault Revelation Effectiveness}
\label{sec:rq3}

\begin{figure*}[t]
\centering

\begin{minipage}{0.23\textwidth}
\centering
\includegraphics[width=\linewidth,trim=0.9cm 0cm 0cm 0.85cm,clip]{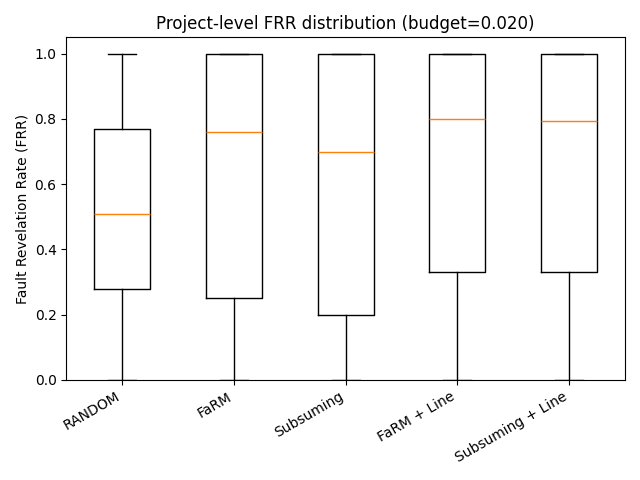}\\
(a) \dataset{Codeflaws} — 2\%
\end{minipage}\hfill
\begin{minipage}{0.23\textwidth}
\centering
\includegraphics[width=\linewidth,trim=0.9cm 0cm 0cm 0.85cm,clip]{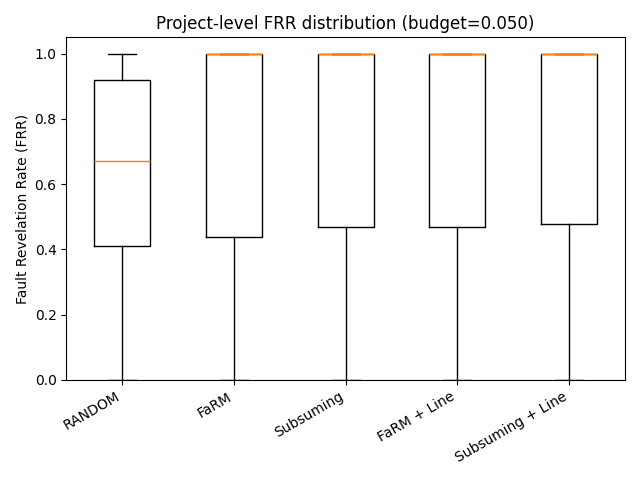}\\
(b) \dataset{Codeflaws} — 5\%
\end{minipage}\hfill
\begin{minipage}{0.23\textwidth}
\centering
\includegraphics[width=\linewidth,trim=0.9cm 0cm 0cm 0.85cm,clip]{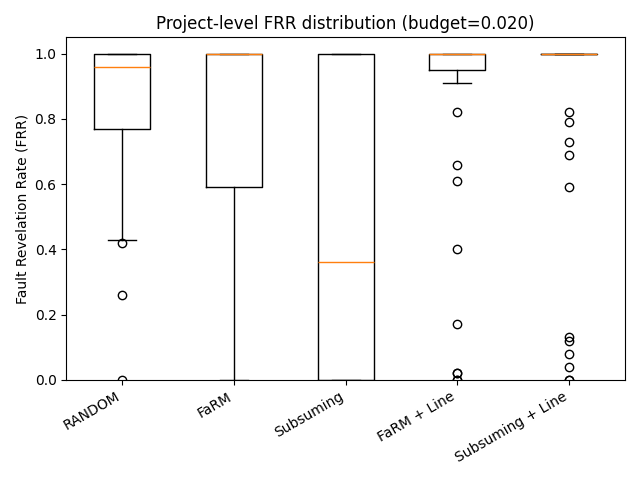}\\
(c) \dataset{CoREBench} — 2\%
\end{minipage}\hfill
\begin{minipage}{0.23\textwidth}
\centering
\includegraphics[width=\linewidth,trim=0.9cm 0cm 0cm 0.85cm,clip]{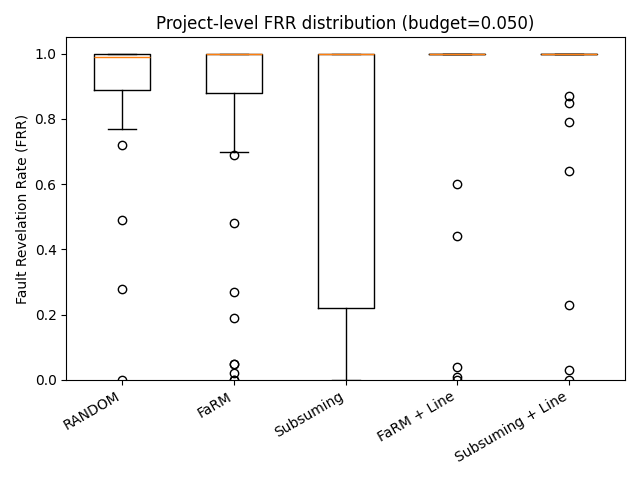}\\
(d) \dataset{CoREBench} — 5\%
\end{minipage}

\caption{Fault Revelation Rate across datasets and selection budgets
(higher is better). Orange lines are medians.
}
\label{fig:rq3_frr}
\end{figure*}

We now examine whether diversification changes \emph{fault revelation
effectiveness} and whether such gains come with additional execution cost.
Figure~\ref{fig:rq3_frr} reports FRR. We compare methods using paired Wilcoxon
signed-rank tests across subjects (programs for \dataset{Codeflaws}, faults for
\dataset{CoREBench}) with Holm correction, and report Vargha--Delaney
$\hat{A}_{12}$ to indicate effect direction. Values $< 0.5$ indicate that
method $B$ tends to outperform method $A$. We further report two
simulation-side cost proxies: the median number of additional mutant
executions (denoted as \emph{median\_cost}) and the median number of sampled
tests until revelation (denoted as \emph{median\_tests}).

\subsubsection{Effectiveness.}
On \dataset{Codeflaws}, at the 2\% budget, learning-based ranking clearly improves (mean) FRR over
\RANDOM (0.524): \FaRM reaches 0.633 and \Sub reaches 0.603, and both
improvements are significant (\RANDOM vs \FaRM: $p_{\mathrm{holm}}=7.96\times10^{-31}$,
$\hat{A}_{12}=0.386$; \RANDOM vs \Sub: $p_{\mathrm{holm}}=4.85\times10^{-20}$,
$\hat{A}_{12}=0.416$). The key observation, however, is what happens once we
\emph{prevent concentration}: pooling pushes FRR higher still (\FaRMLine=0.663,
\SubLine=0.669), and these gains over the corresponding unpooled models are
also significant (\FaRM vs \FaRMLine: $p_{\mathrm{holm}}=1.09\times10^{-10}$,
$\hat{A}_{12}=0.479$; \Sub vs \SubLine: $p_{\mathrm{holm}}=2.67\times10^{-22}$,
$\hat{A}_{12}=0.453$). In other words, at tight budgets, the ranking signal is
valuable, but top-$k$ leaves performance on the table by over-spending the
budget on a few locations. Pooling recovers that value.

At 5\%, the expected \emph{budget compression} appears: all learning-based
variants remain substantially above \RANDOM (0.639), but differences among
them shrink. Both \FaRM (0.727) and \Sub (0.737) significantly outperform
\RANDOM ($p_{\mathrm{holm}}=3.34\times10^{-37}$ and $6.06\times10^{-50}$,
respectively), and pooling yields only small additional separation (e.g.,
\FaRMLine=0.740 is significantly above \FaRM with
$p_{\mathrm{holm}}=6.61\times10^{-5}$, $\hat{A}_{12}=0.491$). As expected, when $k$ grows, multiple strategies ``find'' enough useful
mutants, and the room for improvement contracts.

On \dataset{CoREBench}, the main result is not ``learning helps'' but
``top-$k$ fragility''. At 2\%, unpooled \Sub collapses in FRR (0.492) relative
to \RANDOM (0.836), and the drop is significant (\RANDOM vs \Sub:
$p_{\mathrm{holm}}=2.50\times10^{-4}$, $\hat{A}_{12}=0.642$, i.e., \RANDOM
tends to outperform \Sub). Pooling largely fixes this brittleness: \SubLine
rises to 0.851 and significantly improves over \Sub (\Sub vs \SubLine:
$p_{\mathrm{holm}}=2.50\times10^{-4}$, $\hat{A}_{12}=0.299$). \FaRM shows a
milder version: \FaRM is not significantly different from \RANDOM at 2\%
($p_{\mathrm{holm}}=0.632$), but \FaRMLine increases to 0.861 (not significant
vs \RANDOM under Holm at $n=45$, but directionally higher). At 5\%, the pattern
remains: \Sub is again significantly worse than \RANDOM
($p_{\mathrm{holm}}=2.69\times10^{-2}$), while pooled variants achieve the
highest mean FRR (\FaRMLine=0.911, \SubLine=0.920) and are statistically
indistinguishable from \RANDOM after Holm correction, consistent with limited
power for small effects.

\subsubsection{Blind spots and missed faults.}
To connect these gains to the blind spots identified in RQ1, we partition the \dataset{Codeflaws} subjects based on whether the original top-$k$ selection includes any mutant on the faulty line (at the 2\% budget). The FRR improvement of \LinePool over unpooled top-$k$ ($\Delta\text{FRR}$) is concentrated where top-$k$ misses the faulty line ($+0.066$ for \FaRMLine vs. $+0.004$ when the line is already included).
Moreover, in about 77\% of the trials where \LinePool reveals a fault that top-$k$ misses, the revealing mutant originates from a line absent from the top-$k$ selection.
At the 5\% budget, this trend persists for \FaRM ($+0.054$ $\Delta\text{FRR}$ when top-$k$ misses the faulty line vs. $-0.005$ when included), but is not significant for \Sub due to budget compression.
Conversely, we found no subject in which top-$k$ selection includes the faulty line but \LinePool omits it, at either budget.
On \dataset{CoREBench}, this analysis is not possible, since mutant line numbers cannot be reliably mapped to the fault locations (see Section~\ref{sec:construct}).

\subsubsection{Cost.}
On \dataset{Codeflaws}, the effectiveness gains come with higher median execution cost than \RANDOM,
consistent with learning-based methods steering selection toward mutants whose
revealing tests are found after more rounds. At 2\%, \RANDOM has
\emph{median\_cost} 10 versus 20--23 for learning-based methods (pooling: 22), with
\emph{median\_tests} of 2--3 for all methods; at 5\%, \emph{median\_cost} is 32 for
\RANDOM versus 58--68 for learning-based methods (pooling: 64--66), with
\emph{median\_tests} of 3. Learning-based selection thus trades executions for FRR, while pooling mostly
reallocates \emph{which} mutants are bought with the same budget.

On \dataset{CoREBench}, costs are much larger and vary with the mutant set induced by each
method. At 2\%, \RANDOM has \emph{median\_cost} 1854 and \emph{median\_tests} 7;
\Sub has lower \emph{median\_cost} (1380) yet far worse FRR, illustrating that
``cheaper'' here can simply mean that the process fails to reach revealing tests.
Pooling changes cost in a method-dependent way (\FaRMLine 1635.5, \SubLine
2078; \emph{median\_tests} 7--9). At 5\%, pooled variants are comparable to \RANDOM in cost
(5107.5--5635 vs 5619) while achieving the best mean FRR, whereas \Sub shows high cost (6764.5)
with reduced FRR. Overall, the dominant effect of pooling on
\dataset{CoREBench} is \emph{robustness of
effectiveness}, not cost reduction.

\subsubsection{Worst-case performance analysis.}
To assess performance on difficult cases, we examine the bottom 25\% of
subjects ranked by each baseline's subject-level FRR under the 2\% budget. On
\dataset{Codeflaws}, \FaRMLine outperforms \FaRM in 294 out of 423 cases
($\approx 69\%$), with a mean improvement of $+0.190$ FRR (median $+0.060$).
For \Sub, the gains are even more pronounced: \SubLine improves over the
baseline in 330 out of 423 cases ($\approx 78\%$), with a mean improvement of
$+0.270$ FRR (median $+0.140$). On \dataset{CoREBench}, the improvements are
substantially larger. \FaRMLine outperforms \FaRM in 8 out of 11 cases (mean
$+0.402$, median $+0.350$), while \SubLine improves over \Sub in 9 out of 11
cases with no losses (mean $+0.512$, median $+0.660$). Pooling thus mitigates failure cases rather than only improving average
performance; since the \dataset{CoREBench} bottom quartile contains only 11 faults,
we treat these results as corroborating evidence rather than as a general conclusion.

\begin{tcolorbox}[boxsep=2pt,left=2pt,right=2pt,top=1pt,bottom=1pt]
\noindent\textbf{Answer to RQ3.}
On \dataset{Codeflaws}, learning-based ranking improves FRR over \RANDOM, and
line-level pooling adds gains, especially at the tight 2\% budget, with higher
execution cost than \RANDOM but similar sampled-test counts; these gains are
concentrated where top-$k$ omits the faulty line. On \dataset{CoREBench},
unpooled \Sub performs worse than \RANDOM, consistent with its severe structural collapse and redundancy; pooling
mitigates this brittleness and restores competitive FRR. Cost differences
mainly reflect the selected mutants, not pooling itself.
\end{tcolorbox}

\subsection{RQ4: Design Factors} 
\label{sec:rq4}

\subsubsection{Disentangling Line Pooling and Ranking Signals}
\label{sec:rq4-swsl}

We try to disentangle the respective contributions of \emph{structural diversification} and \emph{ranking quality}. Table~\ref{tab:rq4-ablation-combined} separates the effect of (i) prioritizing \emph{which lines} to draw from ($s_L$) from (ii) prioritizing \emph{which mutants within a line} to draw ($s_W$).

We denote a pooled strategy as \LP{$s_L$}{$s_W$}, where $s_L$ is the signal used to rank source lines and $s_W$ is the signal used to rank mutants within each line. For example, \LP{\FaRM}{\Sub} prioritizes lines according to FaRM scores and selects mutants within each line using Subsuming scores, while \LP{\RandSig}{\RandSig} corresponds to uniform stratified sampling. 
Our main configurations, \FaRMLine and \SubLine (Table~\ref{tab:rq4-ablation-combined}), correspond to \LP{\FaRM}{\FaRM} and \LP{\Sub}{\Sub}, respectively.

\noindent The ablation supports two conclusions:

\paragraph{Pooling contributes the dominant robustness gain.}

Replacing unpooled top-$k$ with any pooled variant collapses Top5Share and increases LineCov, while generally reducing Jaccard across datasets. This indicates that the primary correction is structural: 
pooling expands the effective search space by preventing budget over-allocation to a few dense lines.

\paragraph{Within-line ranking determines who wins once pooling removes collapse.}
Pure stratification \LP{\RandSig}{\RandSig} substantially improves structural diversity but does not achieve the best FRR. In contrast, pooled variants that preserve informative ranking signals (e.g., \LP{\FaRM}{\Sub} or \LP{\Sub}{\FaRM}) consistently achieve higher FRR while exhibiting comparable structural coverage. On \dataset{CoREBench}, \LP{\FaRM}{\Sub} achieves the highest FRR at both budgets. These results suggest that the gains of \LinePool cannot be explained solely by increased spatial coverage. Rather, pooling and ranking play complementary roles: pooling prevents the budget from collapsing onto a few locations, while informative within-line ranking identifies the most useful mutants within the diversified set.

\begin{table*}[t]
\centering
\caption{Ablation results across datasets. FRR and Jaccard are means; others are medians.} 
\vspace{-0.5em}
\small
\setlength{\tabcolsep}{4pt}
\scalebox{0.82}{
\begin{tabular}{llcccc|cccc}
\toprule
Dataset & Method & \multicolumn{4}{c}{Budget 2\%} & \multicolumn{4}{c}{Budget 5\%} \\
\cmidrule(lr){3-6}\cmidrule(lr){7-10}
& & FRR & LineCov & Top5Share & Jaccard &
FRR & LineCov & Top5Share & Jaccard \\
\midrule

\multirow{12}{*}{\dataset{Codeflaws}}
& \RANDOM & 0.524 & 0.396 & 0.797 & 0.366 &
0.639 & 0.547 & 0.723 & 0.366 \\
& \FaRM & 0.633 & 0.286 & 0.944 & 0.302 &
0.727 & 0.466 & 0.857 & 0.285 \\
& \Sub & 0.603 & 0.300 & 0.929 & 0.273 &
0.736 & 0.455 & 0.842 & 0.258 \\
& \FaRMLine & 0.663 & 0.690 & 0.357 & 0.236 &
0.740 & 1.000 & 0.303 & 0.254 \\
& \SubLine & \textbf{0.669} & 0.714 & 0.357 & 0.219 &
0.737 & 1.000 & 0.303 & 0.241 \\
\cmidrule(lr){2-10}
& \LP{\FaRM}{\RandSig} & 0.608 & 0.690 & 0.357 & 0.290 &
0.676 & 1.000 & 0.303 & 0.315 \\
& \LP{\FaRM}{\Sub} & 0.664 & 0.690 & 0.357 & 0.219 &
0.739 & 1.000 & 0.303 & 0.242 \\
& \LP{\Sub}{\FaRM} & 0.668 & 0.714 & 0.357 & 0.235 &
\textbf{0.741} & 1.000 & 0.303 & 0.253 \\
& \LP{\Sub}{\RandSig} & 0.603 & 0.714 & 0.357 & 0.289 &
0.674 & 1.000 & 0.303 & 0.315 \\
& \LP{\RandSig}{\FaRM} & 0.620 & 0.643 & 0.382 & 0.254 &
0.732 & 1.000 & 0.303 & 0.259 \\
& \LP{\RandSig}{\Sub} & 0.620 & 0.643 & 0.382 & 0.243 &
0.731 & 1.000 & 0.303 & 0.247 \\
& \LP{\RandSig}{\RandSig} & 0.560 & 0.650 & 0.381 & 0.311 &
0.668 & 1.000 & 0.305 & 0.318 \\
\midrule

\multirow{12}{*}{\dataset{CoREBench}}
& \RANDOM & 0.836 & 0.299 & 0.143 & 0.264 &
0.904 & 0.471 & 0.126 & 0.262 \\
& \FaRM & 0.746 & 0.073 & 0.601 & 0.335 &
0.822 & 0.172 & 0.415 & 0.278 \\
& \Sub & 0.492 & 0.052 & 0.687 & 0.344 &
0.657 & 0.119 & 0.488 & 0.255 \\
& \FaRMLine & 0.861 & 0.415 & 0.024 & 0.219 &
0.911 & 0.974 & 0.014 & 0.227 \\
& \SubLine & 0.851 & 0.383 & 0.031 & 0.205 &
0.920 & 0.970 & 0.014 & 0.213 \\
\cmidrule(lr){2-10}
& \LP{\FaRM}{\RandSig} & 0.841 & 0.415 & 0.024 & 0.224 &
0.900 & 0.974 & 0.014 & 0.237 \\
& \LP{\FaRM}{\Sub} & \textbf{0.870} & 0.415 & 0.024 & 0.211 &
\textbf{0.925} & 0.974 & 0.014 & 0.219 \\
& \LP{\Sub}{\FaRM} & 0.833 & 0.383 & 0.031 & 0.211 &
0.914 & 0.970 & 0.014 & 0.220 \\
& \LP{\Sub}{\RandSig} & 0.784 & 0.383 & 0.031 & 0.219 &
0.906 & 0.970 & 0.014 & 0.229 \\
& \LP{\RandSig}{\FaRM} & 0.818 & 0.392 & 0.031 & 0.226 &
0.899 & 0.976 & 0.014 & 0.227 \\
& \LP{\RandSig}{\Sub} & 0.837 & 0.392 & 0.031 & 0.220 &
0.912 & 0.976 & 0.014 & 0.221 \\
& \LP{\RandSig}{\RandSig} & 0.795 & 0.391 & 0.031 & 0.235 &
0.893 & 0.976 & 0.014 & 0.238 \\
\bottomrule
\end{tabular}
}
\label{tab:rq4-ablation-combined}
\end{table*}

\subsubsection{Alternative ranking signal sanity check}
\label{sec:rq4-alt-signal}

Line pooling only requires an ordering of mutants and therefore does not depend on how the ranking signal is produced.
To verify that the observed trends are not specific to \FaRM or \Sub, we repeat the selection procedure on \dataset{Codeflaws} using an additional external ranking signal generated by a Large Language Model (LLM)-based heuristic. Specifically, we use the DeepSeek-V3.2 reasoning model to independently score each mutant based solely on static information, including the program source, the mutant's location, and the same static mutant features used by our learning-based models. The model is instructed to estimate the likelihood that a mutant is reachable, propagates observable effects, and is non-redundant with nearby mutants, without access to dynamic information such as tests, executions, or kill sets. The resulting score is used directly to rank mutants for selection.

Table~\ref{tab:rq4-llm-codeflaws} shows the same qualitative patterns as our main results: unpooled top-$k$ selection exhibits strong structural concentration, higher behavioral redundancy, and lower effectiveness, while line pooling notably increases spatial coverage, reduces kill-set overlap, and improves FRR at both budgets.

This experiment serves only as a generality check; we do not interpret the absolute effectiveness of the LLM signal itself.

\begin{table}[t]
\centering
\caption{LLM ablation on \dataset{Codeflaws}; FRR and Jaccard are means; others are medians.}
\vspace{-0.5em}
\small
\setlength{\tabcolsep}{3pt}
\scalebox{0.85}{
\begin{tabular}{llccccc}
\toprule
Budget & Method & FRR & LineCov & MaxShare & Top5Share & Jaccard \\
\midrule
\multirow{2}{*}{2\%}
& LLM        & 0.505 & 0.250 & 0.462 & 1.000 & 0.403 \\
& LLM + Line & 0.555 & 0.764 & 0.067 & 0.333 & 0.352 \\
\midrule
\multirow{2}{*}{5\%}
& LLM        & 0.619 & 0.357 & 0.396 & 0.929 & 0.395 \\
& LLM + Line & 0.669 & 1.000 & 0.062 & 0.303 & 0.338 \\
\bottomrule
\end{tabular}
}
\label{tab:rq4-llm-codeflaws}
\end{table}

\subsubsection{Clustering-based diversification}
\label{sec:rq4-clustering}

Line-level pooling is a deliberately simple and model-agnostic intervention that enforces spatial dispersion by construction. However, it is not the only possible way to reduce redundancy among selected mutants. As an alternative diversification baseline, we evaluate a clustering-based strategy that selects representative mutants from similarity clusters while using the same learning-based ranking signals. Specifically, we apply $k$-means clustering~\cite{Lloyd1982LeastSquaresQuantization} with Euclidean distance over the static mutant features used by the learning-based models, excluding identifiers and the coupling weight, using scikit-learn~\cite{Pedregosa2011Sklearn} with default parameters and K-means++ initialization~\cite{ArthurVassilvitskii2007KMeansPP}, and setting the number of clusters equal to the selection budget $k$. We then select the highest-ranked mutant of each cluster according to the ranking signal.

Since $k$-means depends on its initialization, we repeated the clustering with 10 random seeds on \dataset{Codeflaws} and 3 on \dataset{CoREBench}. The mean FRR varied across seeds by less than 0.01 on \dataset{Codeflaws} and 0.02 on \dataset{CoREBench}; we therefore report the mean over seeds.

Table~\ref{tab:clust-ablation-combined} shows that, on \dataset{Codeflaws}, clustering achieves fault revelation effectiveness and kill-set overlap comparable to pooled selection. Notably, this occurs although clustering remains considerably more concentrated than pooled selection, though less so than unpooled top-$k$ (e.g., LineCov of 0.431 vs.\ 0.286 for \FaRM at 2\%). This suggests that full line coverage is not necessary, as long as the budget does not collapse onto a few locations. On \dataset{CoREBench}, clustering matches pooled selection at 2\% but falls behind at 5\%, and yields higher kill-set overlap at both budgets, suggesting that in larger programs, feature similarity is a less reliable proxy for behavioral diversity than spreading selections across source lines.

Overall, line-level pooling offers a simpler intervention that requires no similarity modeling, clustering procedure, or feature engineering beyond the ranking signal.

\begin{table*}[t]
\centering
\caption{Comparison with clustering-based diversification. FRR and Jaccard are means; others are medians.}
\vspace{-0.5em}
\small
\setlength{\tabcolsep}{4pt}
\resizebox{\textwidth}{!}{
\begin{tabular}{llccccc|ccccc}
\toprule
Dataset & Method 
& \multicolumn{5}{c}{Budget 2\%} 
& \multicolumn{5}{c}{Budget 5\%} \\
\cmidrule(lr){3-7}\cmidrule(lr){8-12}
& 
& FRR & Jaccard & LineCov & MaxShare & Top5Share
& FRR & Jaccard & LineCov & MaxShare & Top5Share \\
\midrule

\multirow{4}{*}{\dataset{Codeflaws}}
& Clustering + \FaRM & 0.674 & 0.240 & 0.431 & 0.228 & 0.733 &
0.752 & 0.254 & 0.625 & 0.194 & 0.653 \\
& Clustering + \Sub & 0.665 & 0.218 & 0.445 & 0.220 & 0.717 &
0.750 & 0.246 & 0.630 & 0.193 & 0.651 \\
& \FaRMLine & 0.663 & 0.236 & 0.690 & 0.071 & 0.357 &
0.740 & 0.254 & 1.000 & 0.062 & 0.303 \\
& \SubLine & 0.669 & 0.219 & 0.714 & 0.071 & 0.357 &
0.737 & 0.241 & 1.000 & 0.062 & 0.303 \\

\midrule

\multirow{4}{*}{\dataset{CoREBench}}
& Clustering + \FaRM & 0.865 & 0.266 & 0.290 & 0.041 & 0.159 &
0.883 & 0.269 & 0.442 & 0.035 & 0.133 \\
& Clustering + \Sub & 0.851 & 0.247 & 0.279 & 0.040 & 0.162 &
0.875 & 0.257 & 0.435 & 0.034 & 0.131 \\
& \FaRMLine & 0.861 & 0.219 & 0.415 & 0.005 & 0.024 &
0.911 & 0.227 & 0.974 & 0.003 & 0.014 \\
& \SubLine & 0.851 & 0.205 & 0.383 & 0.006 & 0.031 &
0.920 & 0.213 & 0.970 & 0.003 & 0.014 \\

\bottomrule
\end{tabular}
}
\label{tab:clust-ablation-combined}
\end{table*}

\subsubsection{Alternative diversification mechanisms sanity check}
\label{sec:rq4-alternatives}

Finally, given the simplicity of \LinePool, we examined whether this simplicity hides a substantially more effective alternative. We therefore compared \LinePool with four established diversification methods, as shown in Table~\ref{tab:alternatives}.
The first is \emph{proportional allocation}, which distributes the budget across lines using Sainte-Lagu\"e apportionment~\cite{SainteLague1910}, as in election-based diversification~\cite{DangCroft2012PM2}. 
The second is maximal marginal relevance (MMR)~\cite{CarbonellGoldstein1998MMR}, which we apply in two variants that measure the similarity between mutants either by the distance between their source lines or by the cosine similarity of their static features.
The third is a determinantal point process (DPP)~\cite{KuleszaTaskar2012DPP} with greedy MAP inference~\cite{Chen2018FastGreedyDPP}, 
and the fourth is submodular facility location~\cite{LinBilmes2011Submodular, Nemhauser1978Submodular}. 
We further evaluated $k$-means clustering on different mutant representations (all static features, random feature subsets (50\% and 25\%), and AST features only), 
mutation operators as alternative strata~\cite{Zhang2010OperatorBased}, 
and alternative line-priority rules for \LinePool (\emph{mean} and \emph{median} instead of the \emph{maximum}).

All strategies consume the same ranking signal (\FaRM or \Sub); where a strategy requires a numeric quality, we use the within-subject percentile of the score. MMR uses $\lambda = 0.5$, 
and facility location is optimized with the lazy greedy algorithm~\cite{Minoux1978LazyGreedy}. 
Clustering follows the setup of Section~\ref{sec:rq4-clustering}. 
To keep the computation tractable on large faults, DPP considers only the 30{,}000 top-ranked mutants as candidates, and facility location measures coverage on a random sample of 5{,}000 mutants.

Table~\ref{tab:alternatives} shows that no method consistently dominates across datasets. Alternatives that perform best on one dataset fall behind on the other: proportional allocation obtains the highest FRR on \dataset{CoREBench} but is below \LinePool on \dataset{Codeflaws} at 2\% and covers fewer lines (Section~\ref{sec:rq5} further shows that it is less robust), whereas feature-based MMR is slightly better on \dataset{Codeflaws} (e.g., 0.764 vs.\ 0.737 for \Sub at 5\%) but below \LinePool on \dataset{CoREBench}. Mutation-operator strata perform comparably on \dataset{Codeflaws} but lower on \dataset{CoREBench}, and mean and median line priorities are at most on par with the maximum. Clustering on all static features or on random feature subsets reaches FRR comparable to \LinePool, whereas restricting it to AST features lowers FRR on \dataset{Codeflaws}.

The common factor behind the gains is thus preventing a tight budget from concentrating on a few locations, while the precise diversification mechanism matters considerably less. In this sense, the program structure is a means to obtain diversity rather than a goal in itself: among the evaluated strategies, \LinePool is the simplest, requires no features or parameters, and is the only one that guarantees spatial coverage.

\begin{table*}[t]
\centering
\caption{Comparison with alternative diversification mechanisms. FRR values are means.}
\vspace{-0.5em}
\small
\setlength{\tabcolsep}{4pt}
\resizebox{0.8\textwidth}{!}{
\begin{tabular}{llcccc|cccc}
\toprule
 & & \multicolumn{4}{c}{\dataset{Codeflaws}} & \multicolumn{4}{c}{\dataset{CoREBench}} \\
\cmidrule(lr){3-6}\cmidrule(lr){7-10}
 & & \multicolumn{2}{c}{\FaRM} & \multicolumn{2}{c}{\Sub} & \multicolumn{2}{c}{\FaRM} & \multicolumn{2}{c}{\Sub} \\
Group & Method & 2\% & 5\% & 2\% & 5\% & 2\% & 5\% & 2\% & 5\% \\
\midrule
\multirow{3}{*}{\textit{Reference}} & \RANDOM & 0.524 & 0.639 & 0.524 & 0.639 & 0.836 & 0.904 & 0.836 & 0.904 \\
 & Top-$k$ & 0.633 & 0.727 & 0.603 & 0.736 & 0.746 & 0.822 & 0.492 & 0.657 \\
 & \LinePool (max) & 0.663 & 0.740 & 0.669 & 0.737 & 0.861 & 0.911 & 0.851 & 0.920 \\
\midrule
\multirow{5}{*}{Diversification} & Proportional (Sainte-Lagu\"e) & 0.639 & 0.736 & 0.637 & 0.750 & 0.904 & 0.917 & 0.923 & 0.928 \\
 & MMR (line distance) & 0.655 & 0.742 & 0.664 & 0.753 & 0.858 & 0.906 & 0.842 & 0.896 \\
 & MMR (feature similarity) & 0.663 & 0.756 & 0.674 & 0.764 & 0.855 & 0.862 & 0.822 & 0.833 \\
 & DPP & 0.657 & 0.746 & 0.673 & 0.755 & 0.844 & 0.882 & 0.831 & 0.848 \\
 & Facility location & 0.660 & 0.733 & 0.641 & 0.744 & 0.779 & 0.820 & 0.578 & 0.668 \\
\midrule
\multirow{4}{*}{Clustering} & All static features & 0.674 & 0.752 & 0.665 & 0.750 & 0.865 & 0.883 & 0.851 & 0.875 \\
 & Random 50\% of features & 0.678 & 0.754 & 0.664 & 0.751 & 0.872 & 0.896 & 0.852 & 0.895 \\
 & Random 25\% of features & 0.675 & 0.749 & 0.666 & 0.747 & 0.858 & 0.885 & 0.848 & 0.892 \\
 & AST features only & 0.649 & 0.732 & 0.650 & 0.736 & 0.834 & 0.867 & 0.872 & 0.904 \\
\midrule
Other strata & Mutation operator & 0.665 & 0.742 & 0.658 & 0.737 & 0.830 & 0.877 & 0.770 & 0.813 \\
\midrule
\multirow{2}{*}{Line priority} & \LinePool (mean) & 0.653 & 0.738 & 0.663 & 0.737 & 0.843 & 0.913 & 0.836 & 0.919 \\
 & \LinePool (median) & 0.642 & 0.737 & 0.659 & 0.734 & 0.829 & 0.910 & 0.850 & 0.896 \\
\bottomrule
\end{tabular}
}
\label{tab:alternatives}
\end{table*}

\begin{tcolorbox}[boxsep=2pt,left=2pt,right=2pt,top=1pt,bottom=1pt,before skip=8pt, after skip=6pt]
\noindent\textbf{Answer to RQ4.}
\LinePool's main gains come from enforcing structural coverage: pooling prevents budget collapse onto a few lines, improving diversity and robustness, while within-line ranking determines which pooled variant performs best. These gains hold across alternative ranking signals, and more elaborate diversification mechanisms (e.g., clustering, MMR, or DPP) do not consistently outperform \LinePool, which remains the simplest way to control structural concentration without degrading effectiveness.
\end{tcolorbox}


\subsection{RQ5: Robustness}
\label{sec:rq5}
\subsubsection{Robustness to Ranking Noise.}
\begin{figure}[t]
\centering
\begin{minipage}{0.44\linewidth}
\centering
\includegraphics[
  width=\linewidth,
  trim=0.9cm 0.2cm 0cm 0.80cm,
  clip
]{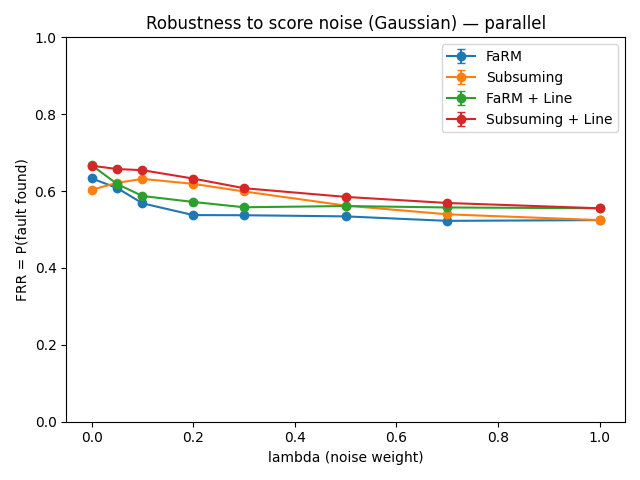}\\[-0.4em]
{\footnotesize (a) \dataset{Codeflaws}}
\end{minipage}
\hfill
\begin{minipage}{0.45\linewidth}
\centering
\includegraphics[
  width=\linewidth,
  trim=0.9cm 0.2cm 0cm 0.80cm,
  clip
]{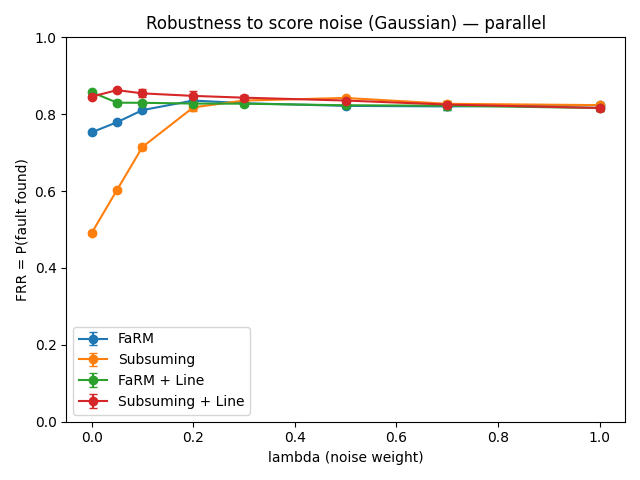}\\[-0.4em]
{\footnotesize (b) \dataset{CoREBench}}
\end{minipage}
\caption{Robustness to Gaussian noise in ranking scores. Curves show mean FRR (y-axis) vs noise magnitude.}
\label{fig:robustness-noise}
\end{figure}

We further evaluate the stability of fault revelation under progressively noisier ranking signals. 
Figure~\ref{fig:robustness-noise} reports FRR as a function of noise magnitude $\lambda$ for \dataset{Codeflaws} and \dataset{CoREBench} at the 2\% budget.

On \dataset{Codeflaws}, increasing noise steadily erodes the advantage of learning-based ranking, with performance gradually approaching that of uninformed selection. 
Across the full noise range, pooled variants (\FaRMLine and \SubLine) remain competitive, indicating that enforcing line coverage stabilizes effectiveness even when within-line ordering is partially corrupted.

On \dataset{CoREBench}, unpooled subsuming ranking shows pronounced brittleness at low noise levels due to severe structural concentration. Moderate noise partially disrupts this collapse, improving performance toward \RANDOM. 
Pooled variants maintain stable and high effectiveness across noise levels, consistent with pooling preventing over-allocation to a few locations.

We applied the same perturbation to proportional allocation, the strongest alternative in Section~\ref{sec:rq4-alternatives}. On \dataset{Codeflaws}, it remains below \LinePool at every noise level, whereas on \dataset{CoREBench} its advantage without noise ($+0.05$ to $+0.07$ FRR) shrinks to at most $+0.03$ once noise is added ($\lambda \geq 0.1$).

Overall, pooling yields more reliable fault revelation under imperfect ranking signals, whereas under extreme noise all methods converge to random selection.

\subsubsection{Out-of-distribution evaluation.}
Table~\ref{tab:rq5-ood} reports results when models are trained on one dataset and evaluated on the other without retraining (Section~\ref{subsub:robustness}).
Across both targets, pooled variants consistently improve both effectiveness and diversity over their unpooled counterparts.
On \dataset{Codeflaws}, \FaRMLine and \SubLine achieve the highest FRR (0.592 and 0.593), while also substantially increasing LineCov and reducing Top5Share and Jaccard.
On \dataset{CoREBench}, the effect is even stronger: unpooled \FaRM degrades markedly below \RANDOM (0.602 vs.\ 0.836), whereas \SubLine achieves the best FRR overall (0.868) together with much stronger spatial dispersion.
Proportional allocation does not share this robustness: on \dataset{Codeflaws}, its FRR falls below that of \LinePool (0.522 and 0.549 vs.\ 0.592 and 0.593), with \FaRM even below unpooled top-$k$, and on \dataset{CoREBench} it is on par with \LinePool for \FaRM and below it for \Sub (0.827 vs.\ 0.868), while exhibiting higher kill-set overlap on both datasets.
These results suggest that line-level pooling improves robustness under distribution shift by reducing over-concentration and limiting the dependence of selection quality on precise score calibration.

\begin{table}[t]
\centering
\caption{Out-of-distribution evaluation. FRR and Jaccard are means; others are medians.}
\vspace{-0.5em}
\small
\setlength{\tabcolsep}{4pt}
\scalebox{0.8}{
\begin{tabular}{llcccc}
\toprule
Dataset & Method & FRR & LineCov & Top5Share & Jaccard \\
\midrule
\multirow{7}{*}{%
  \begin{tabular}[c]{@{}c@{}}
  \dataset{Codeflaws}\\
  (2\%)
  \end{tabular}
}
& \RANDOM   & 0.524 & 0.396 & 0.797 & 0.366 \\
& \FaRM     & 0.545 & 0.342 & 0.875 & 0.387 \\
& \Sub      & 0.514 & 0.240 & 1.000 & 0.401 \\
& \FaRMLine & 0.592 & 0.615 & 0.385 & 0.311 \\
& \SubLine  & 0.593 & 0.632 & 0.385 & 0.295 \\
& Proportional (\FaRM) & 0.522 & 0.457 & 0.700 & 0.376 \\
& Proportional (\Sub)  & 0.549 & 0.444 & 0.733 & 0.362 \\
\midrule
\multirow{7}{*}{%
  \begin{tabular}[c]{@{}c@{}}
  \dataset{CoREBench}\\
  (2\%)
  \end{tabular}
}
& \RANDOM   & 0.836 & 0.299 & 0.143 & 0.264 \\
& \FaRM     & 0.602 & 0.120 & 0.343 & 0.302 \\
& \Sub      & 0.740 & 0.133 & 0.388 & 0.352 \\
& \FaRMLine & 0.741 & 0.393 & 0.033 & 0.252 \\
& \SubLine  & 0.868 & 0.467 & 0.028 & 0.289 \\
& Proportional (\FaRM) & 0.747 & 0.361 & 0.111 & 0.292 \\
& Proportional (\Sub)  & 0.827 & 0.358 & 0.106 & 0.325 \\
\bottomrule
\end{tabular}
}
\label{tab:rq5-ood}
\end{table}

\begin{tcolorbox}[boxsep=2pt,left=2pt,right=2pt,top=1pt,bottom=1pt,before skip=8pt, after skip=8pt]
\noindent\textbf{Answer to RQ5.}
Line-level pooling improves robustness under both noisy rankings and distribution shift. Under Gaussian noise, pooled variants remain more stable and degrade more gracefully as ranking quality deteriorates. Under out-of-distribution evaluation, they consistently outperform their unpooled counterparts on both datasets while preserving stronger structural and behavioral diversity. They are also more robust than proportional allocation, the strongest alternative from RQ4. Overall, pooling reduces sensitivity to score calibration and mitigates brittle concentration under shift.
\end{tcolorbox}

\vspace{-0.5em}

\section{Threats to Validity}
\label{sec:threats}

\subsection{Internal Validity}
Our results depend on the ranking signals used for mutant selection. For
\FaRM, we use the released implementation and treat its output strictly as a
ranking score. For \Sub, we follow Cerebro's feature-based baseline with
10-fold cross-validation, as its neural model is incompatible with our LLVM
bitcode--level mutant representation.
We also include an LLM-generated ranking signal as a sanity check to assess
whether the observed trends are specific to the studied models. Different
models or prompts may produce different rankings. Our goal is only to show
that the structural and behavioral effects of top-$k$ ranking also appear
under an alternative signal.
For the alternative diversification mechanisms, we used standard settings
without tuning (e.g., $\lambda = 0.5$ for MMR) and approximations for DPP and
facility location on large \dataset{CoREBench} faults. Tuned variants might
perform differently. However, our goal was to test whether a substantially
more effective alternative exists, not to optimize each mechanism.

\subsection{External Validity}
\dataset{Codeflaws} contains small competition programs, so the observed
effects may differ in larger systems. \dataset{CoREBench} partially mitigates
this limitation by including real regression faults from large C projects.
Notably, the effects we observe are generally stronger on \dataset{CoREBench}.
For example, the gap in line coverage between top-$k$ selection and \RANDOM is
larger, and unpooled top-$k$ selection falls below \RANDOM in fault revelation
only on this dataset (Table~\ref{tab:rq4-ablation-combined}). This suggests
that the small size of \dataset{Codeflaws} programs is more likely to lead us
to underestimate, rather than inflate, the phenomenon studied here.
Nevertheless, our study is limited to C programs and the mutation operators
and tooling used in these benchmarks. Replication on other languages and
mutation systems would strengthen generality.

\subsection{Construct Validity}
\label{sec:construct}
We measure structural diversity using line-level metrics (LineCov, MaxShare,
and Top5Share). These capture spatial concentration at \LinePool's target
granularity, but not higher-level structures such as functions or modules. We
therefore complement them with faulty-line coverage, behavioral redundancy, and fault-revelation analyses. The faulty-line analysis is limited to \dataset{Codeflaws}. In \dataset{CoREBench}, mutant line numbers refer to an intermediate build representation of the multi-file programs rather than to the original source files, so they cannot be reliably mapped to the known fault locations. This does not affect our other \dataset{CoREBench} results, since \LinePool and the structural metrics only require consistent line identifiers within a subject. Moreover, the analysis is a mechanism check rather than primary evidence: on \dataset{CoREBench}, top-$k$ selection is even more concentrated (LineCov of 0.073 and 0.052 vs.\ 0.299 for \RANDOM at 2\%), and pooling improves FRR in the same direction.

Behavioral redundancy is measured using Jaccard similarity over kill sets,
which depends on the available test suites and the treatment of mutants with
empty kill sets. To mitigate test-suite dependence, both benchmarks supplement
developer-written tests with automatically generated tests under generous time
budgets. To mitigate empty-kill-set sensitivity, we repeat the analysis using
three treatments: similarity 1, similarity 0, and exclusion. The qualitative
conclusions remain unchanged, suggesting that the observed redundancy
reduction is not an artifact of a particular Jaccard formulation.

FRR is a simulation-based proxy for fault revelation and may not capture all
practical cost trade-offs. Therefore, we also report execution-based cost and
the number of selected tests until fault revelation.

\subsection{Conclusion Validity}
FRR evaluation is simulation-based and stochastic. We mitigate variance by
repeating simulations 100 times per subject and using paired statistical
comparisons. \dataset{CoREBench} contains only 45 faults, which limits
statistical power: several differences, such as \FaRMLine versus \RANDOM at
the 2\% budget, are directionally consistent with our findings but not
significant after Holm correction. We therefore treat \dataset{CoREBench}
results as corroborating evidence for the direction of the effects rather
than as precise estimates of their magnitude. To support this interpretation,
we report effect sizes ($\hat{A}_{12}$) alongside $p$-values.

\enlargethispage{2\baselineskip}
\section{Conclusion}
\label{sec:conclusion}

Learning-based mutant selection has emerged as a promising approach for
reducing the cost of mutation testing. Existing work has primarily evaluated
these techniques through effectiveness metrics, such as fault revelation. Our
results suggest that this perspective is incomplete. Learning-based ranking can concentrate the selection budget on a few locations, leaving blind spots and, particularly in larger programs, producing behaviorally redundant mutant sets, despite achieving competitive effectiveness. On \dataset{Codeflaws}, for example, top-$k$
selection outperforms \RANDOM while omitting the faulty line in about half of
the subjects.

We presented \LinePool, a simple, model-agnostic diversification strategy that
reuses existing ranking signals to prioritize both source locations and
mutants within each location. Without modifying or retraining the underlying
models, \LinePool consistently improves the structural and behavioral
diversity of the selected mutant sets while preserving, and often improving,
fault revelation effectiveness and robustness under noisy rankings and
distribution shift. Comparisons with clustering and established
diversification methods further indicate that the gains stem from preventing
budget concentration rather than from a specific diversification mechanism.
When mechanisms reach similar effectiveness, simplicity matters: \LinePool
requires no features, similarity measure, or parameters, and it is the only
evaluated strategy that guarantees spatial coverage.

\LinePool is also easy to adopt in practice, as it requires only a score for
each mutant and its source line. MART, which already computes \FaRM
scores~\cite{ChekamEtAl2019MART}, can apply it directly as a post-processing
step, and we expect it to integrate equally easily with language-model-based
approaches such as $\mu$BERT~\cite{DegiovanniPapadakis2022muBERT} and,
potentially, with other predictive mutation-testing techniques. Limiting the
number of mutants per line is also not new in practice: Google's production
mutation-testing system~\cite{Petrovic2022PracticalScale} and recent LLM-based
test generation~\cite{Chang2026AdverTest} both constrain mutants at the line
level, but without analyzing how such limits affect redundancy or
effectiveness. Our results provide this rationale, showing that beyond
reducing cost, spreading mutants across lines reduces redundancy and, under
tight budgets, improves fault revelation.

More broadly, our findings suggest that better prediction of a specific
mutant property does not necessarily translate into better downstream
outcomes. A ranker that identifies likely subsuming or fault-revealing
mutants can still yield a concentrated selection that reveals fewer faults
than \RANDOM, as \Sub does on \dataset{CoREBench}, while \LinePool improves
fault revelation without changing the underlying predictor. How predictions
are consumed under a budget therefore matters as much as their accuracy,
which is likely to hold for other downstream uses of selected mutants, such
as test generation. Future learning-based mutant selection techniques should
thus be designed and evaluated not only by the effectiveness of the selected
mutants, but also by their behavioral diversity and by whether they leave parts of the program without any selected mutant. Future work
will focus on diversity-aware learning objectives, the explicit incorporation
of diversity into model training, and the evaluation of such approaches
across additional programming languages and mutation systems.

\section*{Data Availability}
Code and data are available at: https://zenodo.org/records/22998124.

\clearpage
\bibliographystyle{ACM-Reference-Format}  
\bibliography{references}

@inproceedings{Petrovic2018Google,
  author    = {Petrovi{\'c}, Goran and Ivankovi{\'c}, Marko},
  title     = {State of Mutation Testing at {Google}},
  booktitle = {Proceedings of the 40th International Conference on Software
               Engineering: Software Engineering in Practice (ICSE-SEIP)},
  year      = {2018}
}

@inproceedings{Ma2021MuDelta,
  author    = {Ma, Wei and Titcheu Chekam, Thierry and Papadakis, Mike
               and Harman, Mark},
  title     = {{MuDelta}: Delta-Oriented Mutation Testing at Commit Time},
  booktitle = {Proceedings of the 43rd IEEE/ACM International Conference on
               Software Engineering (ICSE)},
  year      = {2021}
}

@article{Chang2026AdverTest,
  author  = {Chang, Pengyu and others},
  title   = {Test vs Mutant: Adversarial {LLM} Agents for Robust Unit Test
             Generation},
  journal = {arXiv preprint arXiv:2602.08146},
  year    = {2026}
}

@article{Friedman2002StochasticGB,
  author  = {Friedman, Jerome H.},
  title   = {Stochastic Gradient Boosting},
  journal = {Computational Statistics \& Data Analysis},
  volume  = {38},
  number  = {4},
  pages   = {367--378},
  year    = {2002},
}

@article{Lloyd1982LeastSquaresQuantization,
  author  = {Lloyd, Stuart},
  title   = {Least Squares Quantization in {PCM}},
  journal = {IEEE Transactions on Information Theory},
  volume  = {28},
  number  = {2},
  pages   = {129--137},
  year    = {1982},
}

@inproceedings{cadar2008klee,
  author    = {Cristian Cadar and Daniel Dunbar and Dawson R. Engler},
  title     = {KLEE: Unassisted and Automatic Generation of High-Coverage Tests for Complex Systems Programs},
  booktitle = {Proceedings of the 8th USENIX Symposium on Operating Systems Design and Implementation (OSDI ’08)},
  pages     = {209--224},
  year      = {2008},
  publisher = {USENIX Association},
  address   = {San Diego, CA, USA}
}

@inproceedings{hendrycks2019robustness,
  author    = {Hendrycks, Dan and Dietterich, Thomas},
  title     = {Benchmarking Neural Network Robustness to Common Corruptions and Perturbations},
  booktitle = {International Conference on Learning Representations (ICLR)},
  year      = {2019}
}

@article{Holm1979,
  author  = {Holm, Sture},
  title   = {A Simple Sequentially Rejective Multiple Test Procedure},
  journal = {Scandinavian Journal of Statistics},
  volume  = {6},
  number  = {2},
  pages   = {65--70},
  year    = {1979}
}

@article{Wilcoxon1945,
  author  = {Wilcoxon, Frank},
  title   = {Individual Comparisons by Ranking Methods},
  journal = {Biometrics Bulletin},
  volume  = {1},
  number  = {6},
  pages   = {80--83},
  year    = {1945},
}

@article{Jaccard1901,
  author  = {Jaccard, Paul},
  title   = {\'Etude Comparative de la Distribution Florale dans une Portion des Alpes et des Jura},
  journal = {Bulletin de la Soci\'et\'e Vaudoise des Sciences Naturelles},
  volume  = {37},
  pages   = {547--579},
  year    = {1901},
}

@article{ChekamEtAl2021StubbornMutants,
  author  = {Titcheu Chekam, Thierry and Papadakis, Mike and Cordy, Maxime and Le Traon, Yves},
  title   = {Killing Stubborn Mutants with Symbolic Execution},
  journal = {ACM Transactions on Software Engineering and Methodology},
  volume  = {30},
  number  = {2},
  pages   = {1--23},
  year    = {2021},
}

@inproceedings{ChekamEtAl2019MART,
  author    = {Titcheu Chekam, Thierry and Papadakis, Mike and Le Traon, Yves},
  title     = {{MART}: A Mutant Generation Tool for {LLVM}},
  booktitle = {Proceedings of the 27th ACM Joint European Software Engineering Conference and Symposium on the Foundations of Software Engineering (ESEC/FSE)},
  pages     = {1080--1084},
  year      = {2019}
}

@article{TitcheuChekamEtAl2020SelectingFaultRevealing,
  author  = {Titcheu Chekam, Thierry and Papadakis, Mike and Bissyand{\'e}, Tegawend{\'e} F. and Le Traon, Yves and Sen, Koushik},
  title   = {Selecting Fault Revealing Mutants},
  journal = {Empirical Software Engineering},
  volume  = {25},
  number  = {1},
  pages   = {434--487},
  year    = {2020},
}

@article{GargEtAl2023Cerebro,
  author  = {Garg, Akshaya and Ojdanic, Milan and Degiovanni, Renzo and Titcheu Chekam, Thierry and Papadakis, Mike and Le Traon, Yves},
  title   = {{Cerebro}: Static Subsuming Mutant Selection},
  journal = {IEEE Transactions on Software Engineering},
  volume  = {49},
  number  = {1},
  pages   = {24--43},
  year    = {2023},
}

@inproceedings{GargEtAl2024Vul4JLLMMutants,
  author    = {Garg, Akshaya and Degiovanni, Renzo and Papadakis, Mike and Le Traon, Yves},
  title     = {On the Coupling between Vulnerabilities and {LLM}-Generated Mutants: A Study on {Vul4J} Dataset},
  booktitle = {IEEE International Conference on Software Testing, Verification and Validation (ICST)},
  pages     = {305--316},
  year      = {2024}
}

@inproceedings{GaySalahirad2023CoupledOperators,
  author    = {Gay, Gregory and Salahirad, Amir},
  title     = {How Closely Are Common Mutation Operators Coupled to Real Faults?},
  booktitle = {IEEE International Conference on Software Testing, Verification and Validation (ICST)},
  year      = {2023}
}

@article{GongEtAl2017DominanceWeakMutation,
  author  = {Gong, D. and Zhang, G. and Yao, X. and Meng, F.},
  title   = {Mutant Reduction Based on Dominance Relation for Weak Mutation Testing},
  journal = {Information and Software Technology},
  volume  = {81},
  pages   = {82--96},
  year    = {2017},
}

@inproceedings{GopinathEtAl2016LimitsReduction,
  author    = {Gopinath, Rahul and Alipour, Mohammad Amin and Ahmed, Iftekhar and Jensen, Carlos and Groce, Alex},
  title     = {On the Limits of Mutation Reduction Strategies},
  booktitle = {Proceedings of the 38th International Conference on Software Engineering (ICSE)},
  year      = {2016}
}

@article{LaurentEtAl2022Defects4J2Coupling,
  author    = {Laurent, Thomas and Gaffney, Stephen and Ventresque, Anthony},
  title     = {Re-visiting the Coupling between Mutants and Real Faults with {Defects4J} 2.0},
  booktitle = {Proceedings (Workshop) — Mutation 2022},
  pages     = {189},
  year      = {2022}
}

@inproceedings{PapadakisEtAl2018ICSECorrelation,
  author    = {Papadakis, Mike and Shin, Donghwan and Yoo, Shin and Bae, Doo-Hwan},
  title     = {Are Mutation Scores Correlated with Real Fault Detection? A Large Scale Empirical Study on the Relationship between Mutants and Real Faults},
  booktitle = {Proceedings of the 40th International Conference on Software Engineering (ICSE)},
  pages     = {537--548},
  year      = {2018}
}

@incollection{PapadakisEtAl2019AdvancesInComputersSurvey,
  author    = {Papadakis, Mike and Kintis, Marinos and Zhang, Jie and Jia, Yue and Le Traon, Yves and Harman, Mark},
  title     = {Mutation Testing Advances: An Analysis and Survey},
  booktitle = {Advances in Computers},
  pages     = {275--378},
  publisher = {Elsevier},
  year      = {2019}
}

@article{PizzoletoEtAl2019SLRMutationCost,
  author  = {Pizzoleto, Alessandro V. and Ferrari, Fabiano C. and Offutt, Jeff and Fernandes, L{\'u}cia and Ribeiro, M{\'a}rcio},
  title   = {A Systematic Literature Review of Techniques and Metrics to Reduce the Cost of Mutation Testing},
  journal = {Journal of Systems and Software},
  volume  = {157},
  pages   = {110388},
  year    = {2019},
}

@article{RostamiJalili2023FrMi,
  author  = {Rostami, Taha and Jalili, Saeed},
  title   = {{FrMi}: Fault-Revealing Mutant Identification using Killability Severity},
  journal = {Information and Software Technology},
  volume  = {164},
  pages   = {107307},
  year    = {2023},
}

@article{SunEtAl2017PathAwareReduction,
  author  = {Sun, Chih-An and Xue, Fei and Liu, Hao and Zhang, Xiangyu},
  title   = {A Path-Aware Approach to Mutant Reduction in Mutation Testing},
  journal = {Information and Software Technology},
  volume  = {81},
  pages   = {65--81},
  year    = {2017},
}

@inproceedings{mathur1991performance,
  author    = {Mathur, Aditya P.},
  title     = {Performance, Effectiveness, and Reliability Issues in Software Testing},
  booktitle = {Proceedings of the Fifteenth Annual International Computer Software and Applications Conference (COMPSAC '91)},
  year      = {1991},
  pages     = {604--605},
  address   = {Tokyo, Japan}
}

@article{WongMathur1995CostReduction,
  author  = {Wong, W. Eric and Mathur, Aditya P.},
  title   = {Reducing the Cost of Mutation Testing: An Empirical Study},
  journal = {Journal of Systems and Software},
  volume  = {31},
  number  = {3},
  pages   = {185--196},
  year    = {1995},
}

@inproceedings{OffuttRothermelZapf1993SelectiveMutation,
  author    = {Offutt, A. Jefferson and Rothermel, Gregg and Zapf, Christian},
  title     = {An Experimental Evaluation of Selective Mutation},
  booktitle = {Proceedings of the 15th International Conference on Software Engineering (ICSE)},
  year      = {1993}
}

@misc{FosterEtAl2025MetaMutationGuidedLLM,
  author    = {Harman, Mark and Ritchey, John and Harper, Ian and Sengupta, Shubho and Mao, Ke and Gulati, Aashish and Foster, Cameron and Robert, Henry},
  title     = {Mutation-Guided {LLM}-Based Test Generation at {Meta}},
  booktitle = {Proceedings of the 33rd ACM Joint European Software Engineering Conference and Symposium on the Foundations of Software Engineering (ESEC/FSE 2025)},
  pages     = {180--191},
  year      = {2025},
  publisher = {ACM}
}

@mastersthesis{Hussain2008MutationClustering,
  author = {Hussain, Shamaila},
  title  = {Mutation Clustering},
  school = {Kings College London},
  year   = {2008}
}

@inproceedings{CoreBench2014issta,
author = {B\"{o}hme, Marcel and Roychoudhury, Abhik},
title = {CoREBench: studying complexity of regression errors},
year = {2014},
isbn = {9781450326452},
publisher = {Association for Computing Machinery},
address = {New York, NY, USA},
url = {https://doi.org/10.1145/2610384.2628058},
doi = {10.1145/2610384.2628058},
booktitle = {Proceedings of the 2014 International Symposium on Software Testing and Analysis},
pages = {105–115},
numpages = {11},
location = {San Jose, CA, USA},
series = {ISSTA 2014}
}

@inproceedings{CodeFlaws2017icse-c,
author = {Tan, Shin Hwei and Yi, Jooyong and Yulis and Mechtaev, Sergey and Roychoudhury, Abhik},
title = {Codeflaws: a programming competition benchmark for evaluating automated program repair tools},
year = {2017},
isbn = {9781538615898},
publisher = {IEEE Press},
url = {https://doi.org/10.1109/ICSE-C.2017.76},
doi = {10.1109/ICSE-C.2017.76},
booktitle = {Proceedings of the 39th International Conference on Software Engineering Companion},
pages = {180–182},
numpages = {3},
location = {Buenos Aires, Argentina},
series = {ICSE-C '17}
}

@inproceedings{Ammann2014MinimalMutants,
  author = {Ammann, Paul and Delamaro, Marcio E. and Offutt, Jeff},
  title = {Establishing Theoretical Minimal Sets of Mutants},
  year = {2014},
  publisher = {IEEE Press},
  url = {https://doi.org/10.1109/ICST.2014.14},
  doi = {10.1109/ICST.2014.14},
  booktitle = {Proceedings of the 2014 IEEE Seventh International Conference on Software Testing, Verification and Validation},
  pages = {21--30}
}

@article{OjdanicGKDPT23,
  author       = {Milos Ojdanic and
                  Aayush Garg and
                  Ahmed Khanfir and
                  Renzo Degiovanni and
                  Mike Papadakis and
                  Yves Le Traon},
  title        = {Syntactic Versus Semantic Similarity of Artificial and Real Faults
                  in Mutation Testing Studies},
  journal      = {{IEEE} Trans. Software Eng.},
  volume       = {49},
  number       = {7},
  pages        = {3922--3938},
  year         = {2023},
  url          = {https://doi.org/10.1109/TSE.2023.3277564},
  doi          = {10.1109/TSE.2023.3277564},
}

@inproceedings{CarbonellGoldstein1998MMR,
  author    = {Carbonell, Jaime and Goldstein, Jade},
  title     = {The Use of {MMR}, Diversity-Based Reranking for Reordering Documents and Producing Summaries},
  booktitle = {Proceedings of the 21st Annual International ACM SIGIR Conference on Research and Development in Information Retrieval (SIGIR)},
  pages     = {335--336},
  year      = {1998}
}

@article{KuleszaTaskar2012DPP,
  author  = {Kulesza, Alex and Taskar, Ben},
  title   = {Determinantal Point Processes for Machine Learning},
  journal = {Foundations and Trends in Machine Learning},
  volume  = {5},
  number  = {2--3},
  pages   = {123--286},
  year    = {2012}
}

@inproceedings{Chen2018FastGreedyDPP,
  author    = {Chen, Laming and Zhang, Guoxin and Zhou, Hanning},
  title     = {Fast Greedy {MAP} Inference for Determinantal Point Process to Improve Recommendation Diversity},
  booktitle = {Advances in Neural Information Processing Systems 31 (NeurIPS)},
  year      = {2018}
}

@inproceedings{LinBilmes2011Submodular,
  author    = {Lin, Hui and Bilmes, Jeff},
  title     = {A Class of Submodular Functions for Document Summarization},
  booktitle = {Proceedings of the 49th Annual Meeting of the Association for Computational Linguistics: Human Language Technologies (ACL-HLT)},
  pages     = {510--520},
  year      = {2011}
}

@article{Nemhauser1978Submodular,
  author  = {Nemhauser, George L. and Wolsey, Laurence A. and Fisher, Marshall L.},
  title   = {An Analysis of Approximations for Maximizing Submodular Set Functions---{I}},
  journal = {Mathematical Programming},
  volume  = {14},
  number  = {1},
  pages   = {265--294},
  year    = {1978}
}

@inproceedings{Minoux1978LazyGreedy,
  author    = {Minoux, Michel},
  title     = {Accelerated Greedy Algorithms for Maximizing Submodular Set Functions},
  booktitle = {Optimization Techniques},
  series    = {Lecture Notes in Control and Information Sciences},
  volume    = {7},
  pages     = {234--243},
  publisher = {Springer},
  year      = {1978}
}

@inproceedings{DangCroft2012PM2,
  author    = {Dang, Van and Croft, W. Bruce},
  title     = {Diversity by Proportionality: An Election-Based Approach to Search Result Diversification},
  booktitle = {Proceedings of the 35th International ACM SIGIR Conference on Research and Development in Information Retrieval (SIGIR)},
  pages     = {65--74},
  year      = {2012}
}

@article{SainteLague1910,
  author  = {Sainte-Lagu{\"e}, Andr{\'e}},
  title   = {La repr{\'e}sentation proportionnelle et la m{\'e}thode des moindres carr{\'e}s},
  journal = {Annales scientifiques de l'{\'E}cole Normale Sup{\'e}rieure},
  volume  = {27},
  pages   = {529--542},
  year    = {1910}
}

@inproceedings{ArthurVassilvitskii2007KMeansPP,
  author    = {Arthur, David and Vassilvitskii, Sergei},
  title     = {k-means++: The Advantages of Careful Seeding},
  booktitle = {Proceedings of the 18th Annual ACM-SIAM Symposium on Discrete Algorithms (SODA)},
  pages     = {1027--1035},
  year      = {2007}
}

@article{Pedregosa2011Sklearn,
  author  = {Pedregosa, Fabian and Varoquaux, Ga{\"e}l and Gramfort, Alexandre and Michel, Vincent and Thirion, Bertrand and Grisel, Olivier and Blondel, Mathieu and Prettenhofer, Peter and Weiss, Ron and Dubourg, Vincent and Vanderplas, Jake and Passos, Alexandre and Cournapeau, David and Brucher, Matthieu and Perrot, Matthieu and Duchesnay, {\'E}douard},
  title   = {Scikit-learn: Machine Learning in {P}ython},
  journal = {Journal of Machine Learning Research},
  volume  = {12},
  pages   = {2825--2830},
  year    = {2011}
}

@inproceedings{Zhang2010OperatorBased,
  author    = {Zhang, Lu and Hou, Shan-Shan and Hu, Jun-Jue and Xie, Tao and Mei, Hong},
  title     = {Is Operator-Based Mutant Selection Superior to Random Mutant Selection?},
  booktitle = {Proceedings of the 32nd ACM/IEEE International Conference on Software Engineering (ICSE)},
  pages     = {435--444},
  year      = {2010}
}

@inproceedings{DegiovanniPapadakis2022muBERT,
  author    = {Degiovanni, Renzo and Papadakis, Mike},
  title     = {$\mu${BERT}: Mutation Testing using Pre-Trained Language Models},
  booktitle = {IEEE International Conference on Software Testing, Verification
               and Validation Workshops (ICSTW)},
  year      = {2022}
}

@article{Petrovic2022PracticalScale,
  author  = {Petrovi{\'c}, Goran and Ivankovi{\'c}, Marko and Fraser, Gordon
             and Just, Ren{\'e}},
  title   = {Practical Mutation Testing at Scale: A View from {Google}},
  journal = {IEEE Transactions on Software Engineering},
  year    = {2022}
}

\end{document}